\UseRawInputEncoding
\documentclass[journal,letterpaper]{IEEEtran}
\usepackage{amsmath,amsfonts}
\usepackage[T1]{fontenc}
\usepackage{array}
\usepackage{booktabs}
\usepackage{textcomp}
\usepackage{stfloats}
\usepackage{url}
\usepackage{verbatim}
\usepackage{graphicx}
\usepackage{cite}
\usepackage{color}
\usepackage{mathtools}
\usepackage{multirow}
\usepackage{amssymb}
\usepackage{amsthm}
\usepackage{hhline}
\usepackage{enumitem,calc}
\usepackage{tabularray}
\usepackage[noabbrev]{cleveref}

\newtheorem{remark}{Remark}

\usepackage[symbol]{footmisc}

\allowdisplaybreaks 
\ifCLASSOPTIONonecolumn
  \renewcommand{\baselinestretch}{1.57}

\else
  \renewcommand{\baselinestretch}{0.975}

\fi

\begin{document}

\title{Doppler-Compensated Space-Time Coding for MIMO PMCW Radar}
\author{~Gabriel~Daniel, \IEEEmembership{Student Member,~IEEE},~Hanqing~Wu,~Ashish~Pandharipande, and~Geert~Leus, \IEEEmembership{Fellow,~IEEE}
}
\markboth{IEEE Transactions on Signal Processing}
{Daniel \MakeLowercase{\textit{et al.}}: Doppler-Compensated Space-Time Coding for MIMO PMCW Radar}
\IEEEaftertitletext{\vspace{-1.3\baselineskip}}
\maketitle

\begin{abstract}
Phase modulated continuous wave (PMCW) radar is an attractive choice for automotive sensing due to its high dynamic range, interference robustness, and support for integrated communication and sensing. In multiple-input multiple-output (MIMO) PMCW radar, code division multiplexing (CDM) separates transmit signals, but non-ideal auto- and cross-correlation properties increase range sidelobes and degrade dynamic range. This paper proposes a Doppler-compensated space-time coding (STC)-based MIMO PMCW radar framework that separates transmit signals before range correlation, eliminating cross-correlation terms under perfect compensation and allowing the same binary sequence to be used across transmit antennas. The Doppler influence on STC decoding is analytically characterized, showing that chip-wise and inter-STC-slot phase distortions increase range sidelobes and generate ghost components in the direction-of-arrival (DOA) spectrum. To mitigate these effects, Doppler is estimated first, followed by joint fast-time and STC-slot compensation before decoding. Residual STC leakage under imperfect Doppler estimation is quantified, and the processing is extended to multiple moving targets through Doppler-bin-dependent compensation. Theoretical analysis and simulation results demonstrate reduced range sidelobes and suppression of DOA ghosts, while maintaining flexibility in code sequence design.

\end{abstract}



\IEEEPARstart{P}{hase} modulated continuous wave (PMCW) radar has attracted attention in automotive applications due to its high dynamic range~\cite{White2020_DCM}, interference mitigation~\cite{Bourdoux2017_MI_Radar}, and support for integrated communication and sensing~\cite{Dokhanchi2019_JRC}. A biphase modulated carrier signal is transmitted, with phase changes controlled by a binary sequence. After down-conversion and digitization, correlation and a slow-time discrete Fourier transform (DFT) provide range and velocity estimates, respectively. High dynamic range in range estimation requires a sequence with a thumbtack-like auto-correlation function.

Multiple-input multiple-output (MIMO) radar provides high direction-of-arrival (DOA) resolution through virtual array synthesis~\cite{MIMORADAR}, which requires separability of the transmit signals~\cite{Sun2014_MIMORADAR}. In PMCW radar, this is commonly achieved by code division multiplexing (CDM), where each transmit antenna uses a distinct binary sequence~\cite{Overdevest2020_MIMOPMCW,Antes2022_MIMOPMCW1,Bordoux_MIMOPMCW2}. These sequences should have favorable auto-correlation and low cross-correlation over relevant time delays. However, binary sequences with simultaneously ideal auto- and cross-correlation properties do not exist~\cite{Welch1974_LowerBoundCorr,Peng2002_LowerBoundCorr1}, and non-zero cross-correlation increases range sidelobes~\cite{Overdevest2020_MIMOPMCW}. Sequence families such as $m$-sequences~\cite{Sarwate1980_Mseq} and Heimiller codes~\cite{Heimiller1961_HeimillerCodes} also have limited set sizes and code lengths, while Doppler further degrades their correlation performance. These limitations lead to trade-offs between correlation properties, set size, and code length~\cite{Overdevest2020_MIMOPMCW,Antes2022_MIMOPMCW1,Oliveira2022_MIMOPMCW2,Bauduin2021_CodeDiversity}.

As alternatives to CDM-MIMO, outer coding and time-staggered time division multiplexing (TDM) MIMO for PMCW radar have been introduced \cite{Bordoux_MIMOPMCW2}. In outer-coding-based MIMO, the same binary sequences are used for all transmit antennas, and orthogonality is achieved by multiplying each frame at distinct transmit antennas with the corresponding row of a Hadamard matrix. In time-staggered TDM-MIMO, each antenna starts transmitting signals at different time instants, and the amount of time delay depends on the sequence length. However, these MIMO schemes have drawbacks that reduce their flexibility and estimation performance. For fast-moving targets, applying outer codes over slow time degrades the estimation performance since Doppler phase progression over slow time from moving targets disrupts the orthogonality of the Hadamard matrix  \cite{Sun2014_MIMORADAR}. This Doppler-related problem is not addressed in \cite{Bordoux_MIMOPMCW2}. Furthermore, Hadamard outer coding requires the number of transmit antennas and slow-time indices to be a multiple of $4$ \cite{Bordoux_MIMOPMCW2}. Meanwhile, time-staggered TDM requires the sequence length to increase proportionally with the number of transmit antennas to maintain a desired maximum unambiguous range, while also introducing errors in DOA estimation~\cite{Overdevest2020_MIMOPMCW}. The code diversity technique introduced in \cite{Bauduin2021_CodeDiversity} can also be used to combat high range sidelobes in the CDM-MIMO scheme by using different sequences over different transmit antennas and slow-time slots. The diversification of sequences randomizes the sidelobes, thus preventing them to be constructively added. However, the code diversity technique requires a family of codes with large set size since the number of sequences needed is proportional to the number of transmit antennas and slow-time slots.

In \cite{Xu2019_DopplerPMCW}, a modified order of received signal processing in PMCW radar is presented to remedy the shifted correlation caused by the Doppler effect. Unlike the standard approach, the slow-time DFT is applied first to the received signal samples, and the correlation is then performed on the Doppler-compensated samples to produce the range profile of targets. By using this technique, the sidelobes of the range profile are suppressed, provided that a code sequence having a thumbtack-like auto-correlation function is selected. However, this technique does not solve the non-zero cross-correlation and limitation of code set size. 

In this work, we propose a new space-time coding (STC) approach for MIMO PMCW radar, motivated by~\cite{Song2010}, which introduced an STC design for MIMO pulse radar, and a preliminary version of this work was presented in~\cite{Daniel2026Asilomar}. Our STC framework aims to mitigate high range sidelobes caused by non-zero cross-correlation values of the binary sequences. In particular, signals from different transmit antennas are first encoded by an STC matrix, and a decoding process is performed at the receiver prior to range, velocity, and DOA processing. Unlike \cite{Song2010}, we consider the Doppler effect within chip and accumulation period in our signal model due to moving targets. We analytically show that the Doppler effect destroys the orthogonality of the STC matrix and causes ambiguity in the DOA spectrum. Furthermore, the STC framework does not solve the chip-wise Doppler distortion which reduces the auto-correlation performance of the code. These issues further motivate us to propose a Doppler-compensated STC-based processing algorithm. By modifying the processing order as in \cite{Xu2019_DopplerPMCW}, we first perform velocity processing to obtain the Doppler information, and it is used to correct the Doppler shift at the receiver. By using theoretical and simulation results, we will show that the proposed STC framework with Doppler compensation effectively eliminates cross-correlation between transmitted signals and maintains the auto-correlation performance of the code, without resulting in DOA ambiguity. In addition, the proposed scheme simplifies code sequence design since the same binary sequences can be used across all transmit antennas. Therefore, the MIMO PMCW radar employing our proposed technique provides flexibility in the number of transmit/receive antennas and code design (i.e., correlation properties, set size, and code length), while having an excellent range sidelobe performance.

The main contributions of this paper can be summarized as follows:
\begin{itemize}
    \item An analytical expression of range sidelobes caused by non-ideal auto- and cross-correlation functions is provided, and we present a Doppler-aware signal model for STC-based MIMO PMCW radar capturing chip-wise and repetition-slot phase distortions.
    \item We show that the Doppler effect causes DOA ambiguity and high range sidelobes in the STC-based MIMO PMCW radar, degrading its estimation performance.
    \item A Doppler-compensated STC-based processing is proposed to effectively eliminate high range sidelobes caused by non-ideal cross-correlation, while retaining code's auto-correlation performance and suppressing the DOA ambiguity.    
    \item The proposed technique is extended to multiple-target implementation using Doppler-bin-dependent processing, in which each detected Doppler component is independently compensated, decoded, and processed for range and DOA estimation.
\end{itemize}

\section{Signal Model and Processing for CDM-MIMO PMCW Radar}
\label{sec:signal_model}
\subsection{Transmitted Signal Model}
\label{sec:TX_signal_model}
\begin{figure}[!t]
\centering
\includegraphics[width=1\columnwidth]{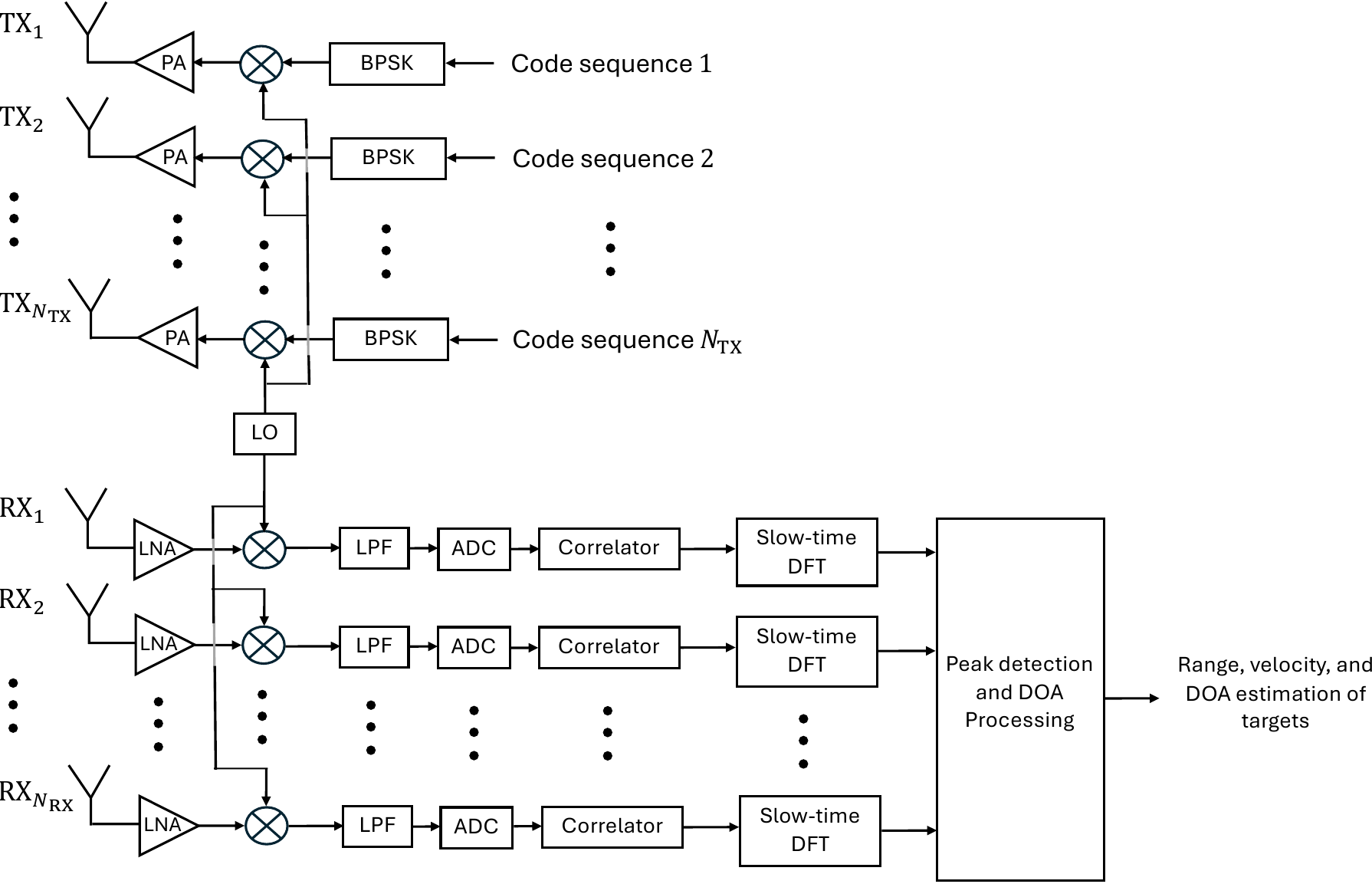}
\caption{System model of CDM-MIMO PMCW radar.}
\label{fig:PMCW_diagram}
\end{figure}

Fig. \ref{fig:PMCW_diagram} shows the CDM-MIMO PMCW radar system with $N_{\mathrm{TX}}$ transmit and $N_{\mathrm{RX}}$ receive antennas. Each transmit antenna sends binary phase-shift keying (BPSK) signals with duration $T_d=NT_c$, in which $N$ chips of duration $T_c$ control the carrier phase. The phase changes are controlled by a code sequence of length $N$. If the code sequences are chosen based on their periodic auto-correlation functions (PACFs) and periodic cross-correlation functions (PCCFs), then these sequences need to be repeated at least once to maintain periodicity, and more repetitions can be implemented to increase the signal-to-noise ratio (SNR) at the receiver. The resulting $L$ repetitions of BPSK codes are then used for range estimation. We refer to a block of $L$ repetitions as a subframe in this paper. To capture velocity information, $M$ subframes are required to create slow-time slots capable of capturing different Doppler frequency shifts. Thus, a full PMCW frame measured at the receiver has duration $T_f=MT_r$, where $T_r=LT_d$ denotes the subframe duration.

In particular, the PMCW signal generated at the $i$-th transmit antenna ($i \in \{0, \dots, N_{\mathrm{TX}}-1 \}$), for a single frame, is
\begin{equation}
        x_i(t) =  \sum_{m=0}^{M-1}\sum_{l=0}^{L-1}\sum_{n=0}^{N-1}c_{i} [n] \mathrm{rect}(t-nT_c-lT_d-mT_r),
    \label{eq:PMCW_signal_model}
\end{equation}
where the BPSK sequence $c_{i}[n]\in\{-1,+1\}$ is set differently for every transmit antenna $i$, and the rectangular function
\begin{equation}
 \mathrm{rect}(t) =\left\{
    \begin{matrix}
    1, &  |t|\leq \frac{T_c}{2}, \\[3pt]
    0, & |t|> \frac{T_c}{2},
    \end{matrix}
    \right.
\end{equation}
is the baseband pulse shaping filter. 
For transmission, the PMCW signal is up-converted to a carrier frequency $f_c$, resulting in
\begin{equation}
    x_{{\rm up},i}(t) = x_i(t) e^{j2\pi f_{c}t}.
    \label{eq:s_up}
\end{equation}

\subsection{Received Signal Model}

\begin{figure*}[!t]
\centering
\begin{minipage}[t]{0.485\textwidth}\vspace{0pt}
\centering
\includegraphics[width=0.830\columnwidth,trim=0 7.75bp 0 8.50bp,clip]{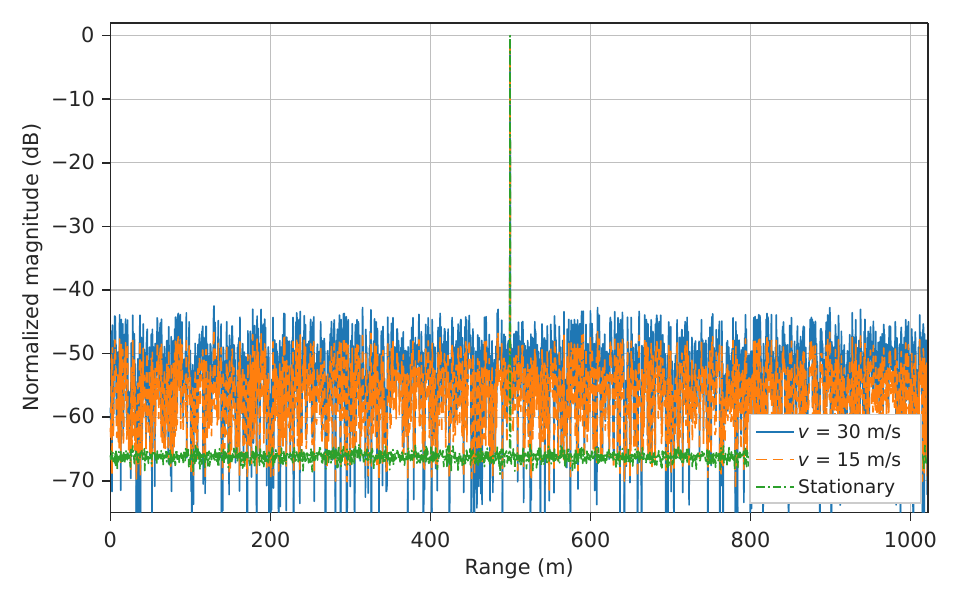}
\caption{The increase of range sidelobes of a target with different values of relative velocity in a SISO setup. The code sequence is an $m$-sequence with length $2047$, and the maximum estimated velocity is around $34$ m/s.}
\label{fig:range_sidelobes_SISO_different_Doppler}
\end{minipage}\hfill%
\begin{minipage}[t]{0.485\textwidth}\vspace{0pt}
\centering
 \includegraphics[width=0.830\columnwidth,trim=0 7.75bp 0 8.50bp,clip]{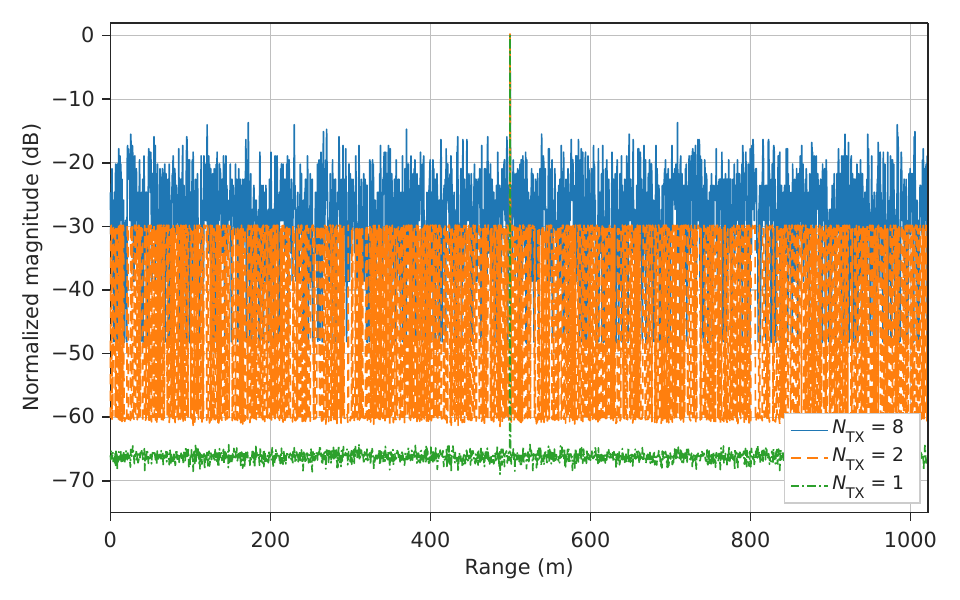}
\caption{The increase of range sidelobes of a target with different numbers of transmit antennas. The code sequence is an $m$-sequence with length $2047$.}
\label{fig:range_sidelobes_different_antennas}
\end{minipage}
\end{figure*}

Let us assume there are $P$ targets, and each target is located at initial range $R_{p}$, moving at radial velocity $v_p$ with respect to the radar transceiver. Then, the time-varying round-trip delay of the $p$-th target is given by
\begin{equation}
\tau_{p}(t) = \frac{2(R_p + v_{p}t)}{c} = \tau_p + \frac{2v_{p}}{c}t,
\label{eq:tau_t}
\end{equation}
where $p=1,2,..., P$ indicates the target index, 
and $c$ represents the speed of light. Therefore, the down-converted signal model at the $j$-th receive antenna ($j \in \{0, \dots, N_{\mathrm{RX}}-1 \}$), for the frame transmitted in \eqref{eq:PMCW_signal_model}, is
\begin{equation}
    y_j(t) = \sum_{p=1}^{P} \alpha_{p}  \phi^{j}_{\mathrm{RX},\theta_{p}} \! \! \sum_{i=0}^{N_{\mathrm{TX}}-1} \! \! \phi^{i}_{\mathrm{TX},\theta_{p}} x_{\rm{up},i}(t - \tau_{p}(t)) e^{-j2\pi f_{c}t} 
\label{eq:s_down}
\end{equation}
where $\alpha_{p}$ denotes the complex amplitude of the $p$-th target, and where $\phi_{\mathrm{RX},\theta_{p}} = e^{j2\pi d_{\mathrm{RX}} \sin(\theta_{p})/\lambda}$ and $\phi_{\mathrm{TX},\theta_{p}} = e^{j2\pi d_{\mathrm{TX}} \sin(\theta_{p})/\lambda}$ are the receive and transmit phase shifts respectively, with $d_{\mathrm{RX}}$ and $d_{\mathrm{TX}}$ denoting the inter-element spacing of the receive and transmit array. These phase shift terms contain the direction-of-arrival (DOA) information of the $p$-th target under a far-field model, denoted as $\theta_{p}$. In this paper, we assume a uniform linear array (ULA) for TX and RX, with $d_{\mathrm{TX}} = \lambda/2$ and $d_{\mathrm{RX}} = N_{\mathrm{TX}} d_{\mathrm{TX}}$ in order to realize a virtual array that is also a ULA with $N_{\mathrm{TX}} N_{\mathrm{RX}}$ antennas at a spacing of $\lambda/2$. Note that we omit the additive noise term in \eqref{eq:s_down} to simplify the mathematical model, yet it is considered in the experiments.

By using \eqref{eq:s_up} and \eqref{eq:tau_t}, we can modify \eqref{eq:s_down} to
\begin{align}
     y_j(t) &= \sum_{p=1}^{P} \alpha_{p} e^{-j2\pi f_{c}\tau_{p}} \phi^{j}_{\mathrm{RX},\theta_{p}}  \! \! \sum_{i=0}^{N_{\mathrm{TX}}-1} \!\! \phi^{i}_{\mathrm{TX},\theta_{p}}  \notag \\ 
     & \hspace{.05\columnwidth} \times x_{i}( (1- 2 v_p/c) t - \tau_p ) e^{-j2\pi f_{c}\frac{2v_{p}}{c}t} \notag \\
     &\approx  \sum_{p=1}^{P} \tilde{\alpha}_{p} \phi^{j}_{\mathrm{RX},\theta_{p}}  \! \! \sum_{i=0}^{N_{\mathrm{TX}}-1} \! \! \phi^{i}_{\mathrm{TX},\theta_{p}} x_i(t - \tau_p)e^{j2\pi f_{D,p}t},
\label{eq:s_down_approx}
\end{align}
where $\tilde{\alpha}_{p}=\alpha_{p}e^{-j2\pi f_{c}\tau_{p}}$ and $f_{D,p}=-f_{c}\frac{2v_{p}}{c}$ is defined as the Doppler frequency shift due to the target's radial velocity. We assume $v_p\ll c$ and negligible range migration, $|v_p|T_f\ll cT_c/2$, to justify the fixed-delay approximation in \eqref{eq:s_down_approx}. 
An analog-to-digital converter (ADC) then samples \eqref{eq:s_down_approx} every $T_c$, which by using~\eqref{eq:PMCW_signal_model} yields
\begin{align}
y_j[m,l,n]= \sum_{p=1}^{P} \tilde{\alpha}_{p} \phi^{j}_{\mathrm{RX},\theta_{p}}  & \! \! \sum_{i=0}^{N_{\mathrm{TX}}-1} \! \! \phi^{i}_{\mathrm{TX},\theta_{p}}  c_{i} [n-\nu_p ] \notag \\
& \times e^{j2\pi f_{D,p}(nT_c+lT_d+mT_r)}. 
\label{eq:s_down_sampled}
\end{align}
Here, $\nu_p=\lfloor\tau_p/T_c\rfloor$ denotes the sequence shift under the range-bin approximation $\tau_p\approx\nu_pT_c$, where $\lfloor x\rfloor$ is the floor function. It is important to note that the delay $\nu_p$ is taken with modulo $N$ due to the $L\geq 2$ repetitions of the BPSK sequence $c_i[n]$. For that same reason, $l=0$ is not considered in~\eqref{eq:s_down_sampled} and only the last $L-1$ repetitions are sampled. This means we have $l=1,2,\dots,L-1$ in~\eqref{eq:s_down_sampled}, with interval $T_d$. On the other hand, the fast-time index $n$ has values $n=0,1,\dots,N-1$, with interval $T_c$, while the slow-time index $m$ satisfies $m=0,1,\dots,M-1$, with interval $T_r$.

\subsection{Standard Processing}
\label{sec:standard_processing}
At the receiver, range processing is performed by correlating \eqref{eq:s_down_sampled} with a replica of the code sequence of the $i^*$-th transmit antenna ($i^* \in \{0, 1, \dots, N_{\mathrm{TX}}-1 \}$), that is 
\begin{equation}
    \begin{split}
        r_{j,i^*}&[m,\nu] = \sum_{l=1}^{L-1}\sum_{n=0}^{N-1} y_j[m,l,n] c_{i^*}[n-\nu] \\
        &= \sum_{p=1}^{P} \tilde{\alpha}_{p} \phi^{j}_{\mathrm{RX},\theta_{p}}  \sum_{l=1}^{L-1}\sum_{i=0}^{N_{\mathrm{TX}}-1}\sum_{n=0}^{N-1} \phi^{i}_{\mathrm{TX},\theta_{p}}  c_{i} [n-\nu_p] \\
        & \hspace{.05\columnwidth} \times c_{i^*}[n-\nu] \ e^{j2\pi f_{D,p}(nT_c+lT_d+mT_r)} \\
        &= \sum_{p=1}^{P} \tilde{\alpha}_{p} \phi^{j}_{\mathrm{RX},\theta_{p}} \sum_{l=1}^{L-1} e^{j2\pi f_{D,p}lT_d} \sum_{i=0}^{N_{\mathrm{TX}}-1} \phi^{i}_{\mathrm{TX},\theta_{p}}   \\
        & \hspace{.05\columnwidth} \times \tilde{r}_{p,i,i^*}[\nu] e^{j2\pi f_{D,p}mT_r}, 
    \end{split}
    \label{eq:range_proc}
\end{equation}
where $\tilde{r}_{p,i,i^*}[\nu]=\sum_{n=0}^{N-1} c_{i} [n-\nu_p]e^{j2\pi f_{D,p}nT_c} c_{i^*}[n-\nu]$, which represents the cross-correlation function of the sequences $c_i[n]$ and $c_{i^*}[n]$, circularly shifted with $\nu_p$ samples and corrupted by the Doppler shift $f_{D,p}$ (if $i^*=i$ we similarly obtain the auto-correlation of $c_{i^*}[n]$).
As mentioned before, the repetition index $l=0$ is not used to ensure circularity in the correlation. Subsequently, a discrete Fourier transform (DFT) is performed on \eqref{eq:range_proc} with respect to the slow-time slots to obtain velocity information, that is
\begin{equation}
    \begin{split}
        \mathcal{R}_{j,i^*}&[k,\nu] = \sum_{m=0}^{M-1} r_{j,i^*}[m,\nu]e^{-j2\pi\frac{m}{M}k} \\
        &= \sum_{p=1}^{P} \tilde{\alpha}_{p} \phi^{j}_{\mathrm{RX},\theta_{p}} \sum_{l=1}^{L-1} e^{j2\pi f_{D,p}lT_d} \sum_{i=0}^{N_{\mathrm{TX}}-1} \phi^{i}_{\mathrm{TX},\theta_{p}}  \\ 
        & \hspace{.05\columnwidth} \times \tilde{r}_{p,i,i^*}[\nu] \sum_{m=0}^{M-1} e^{j2\pi\frac{m}{M} (Mf_{D,p}T_r - k)}.
    \end{split}
    \label{eq:velocity_proc}
\end{equation}

\begin{remark} \label{rmk1}
    The term $e^{j2\pi f_{D,p}nT_c}$ negatively influences the auto- and cross-correlation performance of the code sequence, and we will later on show a technique to solve this issue.
\end{remark}

We call the term $\mathcal{R}_{j,i^*}[k,\nu]$ in \eqref{eq:velocity_proc} a delay-Doppler map (DDM), and a set of $P$ targets can be detected from such a map by computing 
\begin{equation}
    \{(\hat{k}_p,\hat{\nu}_p)\}_{p=1}^{P}= \arg \max_{(k,\nu)}|\mathcal{R}_{j,i^*}[k,\nu]|^2,
\end{equation}
assuming that the true range and velocity of the targets fall exactly at the delay-Doppler bin. Using signed Doppler bins $k\in\{-M/2,\ldots,M/2-1\}$ for even $M$, the estimated range and velocity of each target (denoted as $\hat{R}_p$ and $\hat{v}_p$, respectively) are then given by 
\begin{equation}
        \hat{R}_p = \frac{ \hat{\nu}_p \, c \, T_c}{2} \quad \text{and} \quad
        \hat{v}_p = -\frac{\hat{k}_p \, c}{2MT_rf_c}.
\end{equation}
Here, $\arg\max$ selects the $P$ strongest distinct local peaks. The number of targets is unknown in real-world scenarios, and some adaptive detection techniques such as the constant false alarm rate (CFAR) method \cite{Richards2005_Radar} with its variants can be considered. Furthermore, a final decision on the location of the peaks can also be determined by combining $N_{\mathrm{TX}}  N_{\mathrm{RX}}$ different $\mathcal{R}_{j,i^*}[k,\nu]$ as suggested in \cite{Cagan2026_PMCWFusion}, which improves detection performance in low-SNR scenarios. Lastly, DOA processing is performed by taking a spatial DFT over the $Q=N_{\mathrm{TX}} N_{\mathrm{RX}}$ virtual array indices on the detected $P$ locations. To simplify the presentation of these DOA estimation steps, let us assume that the CDM yields an ideal separation between transmitted signals, i.e.,
\begin{equation}
 \tilde{r}_{p,i,i^*}[\nu]  =\left\{
\begin{matrix}
N\delta[\nu-\nu_p], & \text{for }i=i^*,\\
0, & \text{for } i \neq i^*.
\end{matrix}
\right.          
\label{eq:ideal_sequence}
\end{equation}
This allows us to rewrite \eqref{eq:velocity_proc} as
\begin{align}
        \mathcal{R}_{j,i^*}[k,\nu] 
        &= \sum_{p=1}^{P} \tilde{\alpha}_{p} \sum_{l=1}^{L-1} e^{j2\pi f_{D,p}lT_d} \sum_{m=0}^{M-1} e^{j2\pi\frac{m}{M} (Mf_{D,p}T_r - k)} \notag \\
        & \hspace{.005\columnwidth} \times N\delta[\nu-\nu_p] e^{j\pi \sin(\theta_{p})q}= \mathcal{R}_{q}[k,\nu], \label{eq:velocity_proc_ideal2}
\end{align}
where we used $q = jN_{\mathrm{TX}}+i^* \in \{0, 1, \dots, Q-1 \}$ to indicate the virtual array index. For a target isolated in range and lying on a Doppler DFT bin, the spatial DFT is implemented at its detected location. Specifically, for the $p$-th detected target at $(\hat{k}_p, \hat{\nu}_p)$, we compute
\begin{equation}
    \begin{split}
        & \mathcal{R}_{\omega}[\hat{k}_p,\hat{\nu}_p] = \sum_{q=0}^{Q-1} \mathcal{R}_{q}[\hat{k}_p,\hat{\nu}_p]e^{-j2\pi\frac{q}{Q}\omega} \\ 
       &= \tilde{\alpha}_{p} N M \sum_{l=1}^{L-1} e^{j2\pi f_{D,p}lT_d} \sum_{q=0}^{Q-1} e^{j2\pi \frac{q}{Q}(\frac{Q}{2}\sin(\theta_{p})-\omega)},
    \end{split}
    \label{eq:DOA_spectrum}
\end{equation}
where $\omega$ represents the signed spatial frequency bin. Selecting the peak $\hat{\omega}_p$ of~\eqref{eq:DOA_spectrum} for the $p$-th target, its DOA estimate is obtained as
\begin{equation}
    \hat{\theta}_{p} = \arcsin{\left(\frac{2}{Q}\hat{\omega}_p\right)}.
    \label{eq:DOA_est}
\end{equation}
\begin{remark} \label{rmk2}
    The commonly used code sequences do not have ideal auto- and cross-correlation properties as in~\eqref{eq:ideal_sequence}, and Doppler further degrades these properties. We will show its effect on range and DOA estimation in Section~\ref{sec:Doppler_issues_STC_proc}. In Section~\ref{sec:Doppler_mitigation}, a technique to mitigate this effect is proposed.
\end{remark}

\subsection{Range Sidelobe Analysis}
\label{sec:range_sidelobe_analysis}
In this subsection, we analyze the behavior of the range sidelobes in the CDM-MIMO PMCW radar with non-ideal auto- and cross-correlation caused by the Doppler effect and MIMO configuration. To simplify our analysis, we consider a single target scenario, i.e., $P=1$. First, we look at the performance degradation caused solely by the imperfect auto-correlation due to Doppler. For an on-grid target in a single-input single-output (SISO) scenario, setting $k=\hat{k}_1$ in \eqref{eq:velocity_proc}, the range profile of the target can be expressed as
\begin{equation}
\mathcal{R}[\hat{k}_1,\nu]=\tilde{\alpha}_{1} \sum_{l=1}^{L-1} e^{j2\pi f_{D,1}lT_d}
 M   \tilde{r}_{1}[\nu], 
\label{eq:range_profile1}
\end{equation}
where $\tilde{r}_{1}[\nu]=\sum_{n=0}^{N-1} c [n-\nu_1]e^{j2\pi f_{D,1}nT_c} c[n-\nu]$. Note that we drop the $i$, $i^*$, and $j$ indices due to the SISO case. Let us assume that $c[n]$ possesses ideal auto-correlation as indicated in \eqref{eq:ideal_sequence} and that the target is moving (i.e., $f_{D,1}\neq0$). 
For an unambiguous Doppler frequency interval with respect to the chip-rate sampling $1/T_{c}$, i.e., $-1/(2T_c) \leq f_{D,1} < 1/(2T_c)$,  at $\nu  = \nu_{1}$, we then obtain 
\begin{equation}
\lvert \tilde{r}_{1}[\nu_{1}] \rvert = \left| \sum_{n=0}^{N-1} e^{j2\pi  f_{D,1}nT_c } \right| < N,   
\label{eq:autocorr_upper}
\end{equation}
since any non-zero Doppler frequency yields $f_{D,1}T_{c}\notin\mathbb{Z}$. Furthermore, the orthogonality of $c[n-\nu]$ and $c[n-\nu_{1}]e^{j2\pi  f_{D,1}nT_c }$ cannot be guaranteed $\forall \nu \neq \nu_{1}$, hence $\lvert \tilde{r}_{1}[\nu] \rvert$ will not be in the form of $N \delta[\nu-\nu_1]$ anymore. Thus, the non-zero Doppler degrades the auto-correlation performance of the sequence, and it reduces the peak-to-sidelobe ratio in the range profile of the target. Fig.~\ref{fig:range_sidelobes_SISO_different_Doppler} shows the normalized range cut of a target in a SISO setup. The range sidelobes increase with velocity in the illustrated cases. 

\begin{figure}[!t]
\centering
 \includegraphics[width=0.830\columnwidth,trim=0 7.75bp 0 8.50bp,clip]{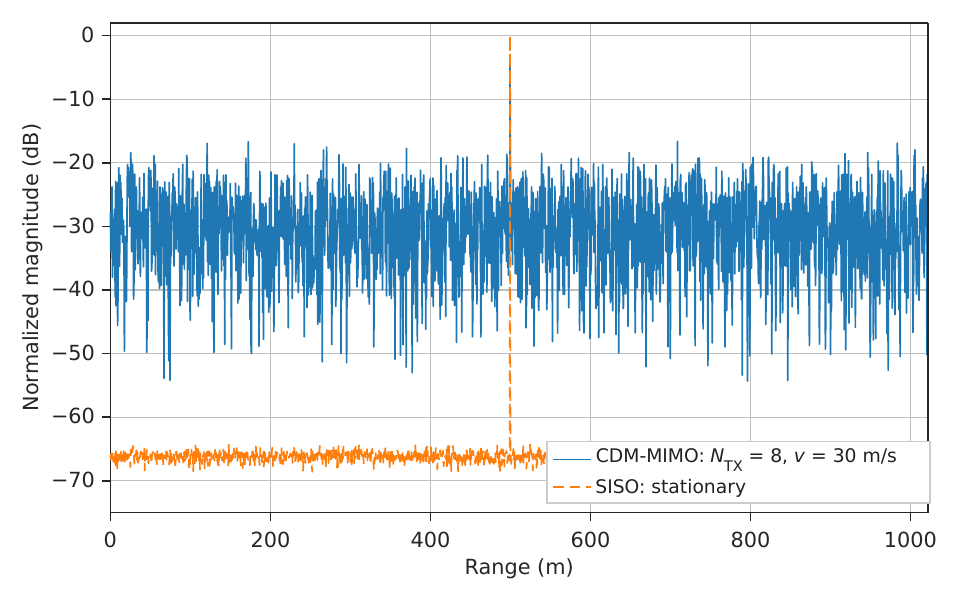}
\caption{Range sidelobe comparison between a stationary target in a SISO setup and a moving target in CDM-MIMO setting. The code sequence is an $m$-sequence with length $2047$.}
\label{fig:range_sidelobes_MIMOvsSISO}
\end{figure}

Now, let us observe the range sidelobes caused only by the non-zero cross-correlation in a MIMO scenario. We assume that the detected target has the same peak location across $Q$ range-velocity maps, thus we can choose an arbitrary $\mathcal{R}_{j,i^*}[k,\nu]$. By setting $f_{D,1}=0$ and $k=\hat{k}_1=0$ in \eqref{eq:velocity_proc}, the range profile of the target can be expressed as
\begin{equation}
\mathcal{R}_{j,i^*}[0,\nu]=\tilde{\alpha}_{1} \phi^{j}_{\mathrm{RX},\theta_{1}} (L-1)
 M \sum_{i=0}^{N_{\mathrm{TX}}-1}  \phi^{i}_{\mathrm{TX},\theta_{1}} \tilde{r}_{1,i,i^*}[\nu].
\label{eq:range_profile2}
\end{equation}
It can be observed from \eqref{eq:range_profile2} that the range profile of a target is a summation of one PACF (when $i = i^*$) and $N_{\mathrm{TX}}-1$ PCCF terms (when $i \neq i^*$). However, as presented in \cite{Overdevest2020_MIMOPMCW}, the commonly used sequences in PMCW radar (e.g., $m$-sequence, Gold, APAS, ZCZ) show relatively high PCCF values, and these values exceed the Welch bound \cite{Welch1974_LowerBoundCorr} for some time delays. Although we have a thumbtack-like PACF, the high PCCF values will contribute to high range sidelobes, and it significantly reduces the dynamic range in the range profile as $N_{\mathrm{TX}}$ increases. Fig. \ref{fig:range_sidelobes_different_antennas} demonstrates the increase of range sidelobes in the CDM-MIMO configuration with higher values of $N_{\mathrm{TX}}$ when the target is stationary. Lastly, in Fig. \ref{fig:range_sidelobes_MIMOvsSISO}, we compare the range sidelobe performance of a single stationary target in a SISO setup with a moving target in a CDM-MIMO setting to show the combined effect from non-ideal auto- and cross-correlation functions.

\begin{figure}[!t]
\centering
 \includegraphics[width=1.000\columnwidth,trim=0 18.25bp 0 9.25bp,clip]{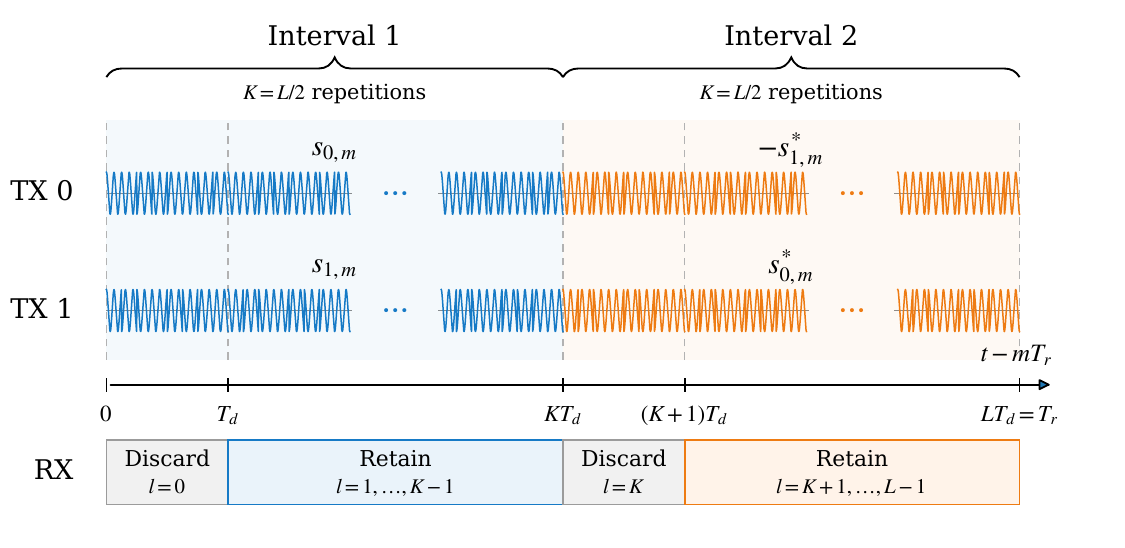}
\caption{Continuous-time STC transmission for two transmit antennas with \(L=2K\) repetitions per subframe, where \(K\geq2\) is an integer. Each coding interval spans \(K\) repetitions. The receiver discards the first repetition of each coding interval, corresponding to \(l=0\) and \(l=K\), and retains the remaining \(L-2\) repetitions for receiver processing.}
\label{fig:STC_transmitted_signal}
\end{figure}

\section{MIMO PMCW Radar with STC-based Processing}
\label{sec:proposed_STC_and_Doppler_Mitigation}
\subsection{STC-based Processing}
In this section, we show how to utilize STC for MIMO PMCW radar. We use the Alamouti STC~\cite{Alamouti1998_Alamouti} to illustrate the technique, hence $N_{\mathrm{TX}}=2$ throughout this section. We also consider a single target to further simplify the signal processing model. The continuous-time
baseband signal transmitted by the $i$-th antenna with STC structure is
\begin{equation}
\begin{aligned}
\widetilde{x}_i(t)&=
\sum_{m=0}^{M-1}
\sum_{a=0}^{1}
\sum_{b=0}^{K-1}
\sum_{n=0}^{N-1}
[\mathbf{X}_m]_{i,a}\,c_i[n] \\
&\quad \times
\operatorname{rect}\!\left(
t-nT_c-(aK+b)T_d-mT_r
\right),
\ i\in\{0,1\}.
\end{aligned}
\label{eq:STC_PMCW_frame}
\end{equation}
Here, $a\in\{0,1\}$ denotes the Alamouti coding interval,
$b\in\{0,\ldots,K-1\}$ denotes the repetition within the coding interval, $L=2K \in \mathbb{Z}$,
$l=aK+b$ is the repetition index within the subframe, and $[\mathbf{X}_m]_{i,a}$ denotes the $(i,a)$-th entry of the STC coding matrix $\mathbf{X}_m$, given by
\begin{equation}
    \mathbf{X}_{m}=
            \begin{bmatrix}
            s_{0,m} & -s_{1,m}^{*} \\
            s_{1,m} & s_{0,m}^{*} 
        \end{bmatrix}.
\label{eq:STC_matrix}
\end{equation}
It is important to note that the repetition within the coding interval is needed to ensure periodicity, and only $L-N_{\mathrm{TX}}$ code repetitions within a subframe will be used for further processing in the receiver as shown in Fig.~\ref{fig:STC_transmitted_signal}.

Since this paper considers only radar-centric applications, the STC symbols can be made uniform across the $M$ subframes, i.e., $\mathbf X_m=\mathbf X$. Assuming that the maximum round-trip delay satisfies $\tau_{\max}<T_d$, we discard the first received repetition of each coding interval, corresponding to $l=0$ and $l=K$. Thus, we can define $\mathcal L_{\mathrm{ret}}$ as a set of retained indices, that is
\begin{equation}
\mathcal L_{\mathrm{ret}}
=\{1,\ldots,K-1\}\cup\{K+1,\ldots,2K-1\}.
\end{equation}
For $a\in\{0,1\}$, we can stack the retained samples at the receiver in a column vector as
\begin{equation}
\mathbf{y}_{j,a}[m,n]
=
\begin{bmatrix}
y_j[m,aK+1,n]\\
\vdots\\
y_j[m,aK+K-1,n] 
\end{bmatrix},
\label{eq:yj_stacked}
\end{equation}
and it is represented compactly as
\begin{equation}
\mathbf y_j^{\mathrm{ret}}[m,n]
=\begin{bmatrix}
\mathbf y_{j,0}^{T}[m,n]&\mathbf y_{j,1}^{T}[m,n]
\end{bmatrix}^{T}. 
\label{eq:yj_compact}
\end{equation}
Now, let us define
\begin{equation}
\begin{aligned}
\mathbf c[n-\nu_1]
&\triangleq\begin{bmatrix}
c_0[n-\nu_1]\\
\phi_{\mathrm{TX},\theta_1}c_1[n-\nu_1]
\end{bmatrix},\\
\mathbf D_K(f_{D,1})
&\triangleq\operatorname{diag}\!\left(
\left\{e^{j2\pi f_{D,1} lT_d}\right\}_{l\in\mathcal L_{\mathrm{ret}}}
\right),
\end{aligned}
\end{equation}
where the diagonal entries of $\mathbf D_K(f_{D,1})$ are arranged in an increasing order of $l$. The values of $\mathbf y_j^{\mathrm{ret}}[m,n]$ are obtained by modifying \eqref{eq:s_down_sampled} based on \eqref{eq:STC_PMCW_frame}, yielding
\begin{equation}
\begin{aligned}
\mathbf y_j^{\mathrm{ret}}[m,n]
={}&\tilde{\alpha}_1\phi_{\mathrm{RX},\theta_1}^{j}
e^{j2\pi f_{D,1}(nT_c+mT_r)}
\\
&\times\mathbf D_K(f_{D,1})
\left(\mathbf X^{T}\otimes\mathbf1_{K-1}\right)
\mathbf c[n-\nu_1],
\end{aligned}
\label{eq:STC_received}
\end{equation}
where $\otimes$ denotes the Kronecker product and $\mathbf1_{K-1}$ is a column vector of ones with $K-1$ entries. 
To simplify the range sidelobe analysis, we first consider a stationary target, i.e., $f_{D,1}=0$, and thus \eqref{eq:STC_received} becomes
\begin{equation}
\mathbf y_j^{\mathrm{ret}}[m,n]
=
\left(\mathbf X^{T}\otimes\mathbf1_{K-1}\right)
\mathbf c[n-\nu_1]\,
\tilde{\alpha}_1\phi_{\mathrm{RX},\theta_1}^{j},
\label{eq:STC_received_rewrite}
\end{equation}
where $T$ denotes a matrix transpose operation.
Assuming the transmit and receive antennas are co-located~\cite{Li2007_MIMORADAR_Colocated}, $\mathbf{X}$ is known at the receiver, and we introduce a pre-processing step that decodes the received samples as follows
\begin{equation}
\begin{split}
\widetilde{\mathbf y}_j[m,n]
&=\left(\mathbf X^{*}\otimes\mathbf I_{K-1}\right)
\mathbf y_j^{\mathrm{ret}}[m,n]\\
&=\left[
\left(\mathbf X^{*}\mathbf X^{T}\right)
\otimes\mathbf1_{K-1}
\right]\mathbf c[n-\nu_1]\\
&\quad\times\tilde{\alpha}_1\phi_{\mathrm{RX},\theta_1}^{j}\\
&=\beta\left(\mathbf c[n-\nu_1]\otimes\mathbf1_{K-1}\right)
\tilde{\alpha}_1\phi_{\mathrm{RX},\theta_1}^{j}\\
&=\begin{bmatrix}
\widetilde{\mathbf y}_{j,0}[m,n]\\
\widetilde{\mathbf y}_{j,1}[m,n]
\end{bmatrix},
\end{split}
\label{eq:STC_decoding}
\end{equation}
where $\mathbf I_{K-1}$ is an identity matrix and $\beta=|s_0|^2+|s_1|^2$. In the last equality of \eqref{eq:STC_decoding}, we use the transmit-antenna index $i$ (instead of the coding-interval index $a$), since the transmitted sequence coming from different transmit antennas can be identified in each entry of $\widetilde{\mathbf y}_j[m,n]$. The decoded samples associated with transmit antenna $i$ are stacked as
\begin{equation}
\widetilde{\mathbf y}_{j,i}[m,n]
=\begin{bmatrix}
\tilde y_j[m,i,1,n]\\
\vdots\\
\tilde y_j[m,i,K-1,n]
\end{bmatrix},\quad i\in\{0,1\}.
\end{equation}
The pre-processing step in \eqref{eq:STC_decoding} is performed for all $M$ subframes. The range processing is then performed as
\begin{equation}
\begin{aligned}
r_{\mathrm{STC},j,i}[m,\nu]
&=\sum_{b=1}^{K-1}\sum_{n=0}^{N-1}
\tilde y_j[m,i,b,n]c_i[n-\nu]\\
&=(K-1)\tilde{\alpha}_1\phi_{\mathrm{RX},\theta_1}^{j}
\beta\phi_{\mathrm{TX},\theta_1}^{i}\tilde r_i[\nu],
\end{aligned}
\label{eq:range_proc_STC}
\end{equation}
where $\tilde r_i[\nu]=\sum_{n=0}^{N-1}c_i[n-\nu_1]c_i[n-\nu]$. Although noise is omitted in the signal model, independent noise contributions with equal variance across repetitions give an SNR gain of $K-1$. Subsequently, applying the slow-time DFT to \eqref{eq:range_proc_STC} yields
\begin{equation}
\begin{aligned}
\mathcal R_{\mathrm{STC},j,i}[k,\nu]
&=\sum_{m=0}^{M-1}r_{\mathrm{STC},j,i}[m,\nu]
e^{-j\frac{2\pi}{M}mk}\\
&=(K-1)\tilde{\alpha}_1\phi_{\mathrm{RX},\theta_1}^{j}
\phi_{\mathrm{TX},\theta_1}^{i}\beta\tilde r_i[\nu]\\
&\quad\times\sum_{m=0}^{M-1}e^{-j\frac{2\pi}{M}mk}.
\end{aligned}
\label{eq:velocity_proc_STC}
\end{equation}
For this stationary target, the signal contribution is located at $k=0$. As in Section~\ref{sec:standard_processing}, we use the virtual array index $q=jN_{\mathrm{TX}}+i$, with $Q=2N_{\mathrm{RX}}$, and denote $\mathcal R_{\mathrm{STC},j,i}[k,\nu]$ by $\mathcal R_{\mathrm{STC},q}[k,\nu]$. The spatial DFT at the true range bin $\nu_1$ is
\begin{equation}
\begin{aligned}
\mathcal R_{\mathrm{STC},\omega}[0,\nu_1]
&=\sum_{q=0}^{Q-1}\mathcal R_{\mathrm{STC},q}[0,\nu_1]
e^{-j2\pi\frac qQ\omega}\\
&=(K-1)\tilde{\alpha}_1\beta NM\\
&\quad\times\sum_{q=0}^{Q-1}
e^{j2\pi\frac qQ\left(\frac Q2\sin\theta_1-\omega\right)}.
\end{aligned}
\label{eq:DOA_proc}
\end{equation}
The DOA estimate can be obtained using \eqref{eq:DOA_est}. At $k=0$, the range profile is
\begin{equation}
\mathcal R_{\mathrm{STC},j,i}[0,\nu]
=(K-1)\tilde{\alpha}_1\phi_{\mathrm{RX},\theta_1}^{j} \beta M\phi_{\mathrm{TX},\theta_1}^{i}\tilde r_i[\nu].
\label{eq:range_profile_STC}
\end{equation}
Thus, the accumulated range response is free of PCCF terms, and its sidelobes are limited by the PACF of the selected sequence. 

Compared to the conventional CDM processing for the same subframe duration, the proposed STC-based processing will have lower peak power since only $L-2$ repetitions are used for the receiver processing. Additionally, the same code sequence can be used across transmit antennas, i.e., $c_0[n]=c_1[n]=\cdots=c_{N_{\mathrm{TX}}-1}[n]$, since \eqref{eq:STC_decoding} separates the transmitted components. As a result, code design is simplified to selecting one sequence with a thumbtack-like PACF. For example, an $m$-sequence~\cite{Sarwate1980_Mseq} has the deterministic PACF
\begin{equation}
\tilde r_i[\nu]
=\sum_{n=0}^{N-1}c_i[n-\nu_1]c_i[n-\nu]
=
\begin{cases}
N,&\nu=\nu_1,\\
-1,&\text{otherwise}.
\end{cases}
\label{eq:mseq_PACF}
\end{equation}

\subsection{Doppler Effect on STC-based MIMO PMCW Radar}
\label{sec:Doppler_issues_STC_proc}
\begin{figure}[!t]
\centering
\includegraphics[width=0.830\columnwidth,trim=0 4.00bp 0 4.75bp,clip]{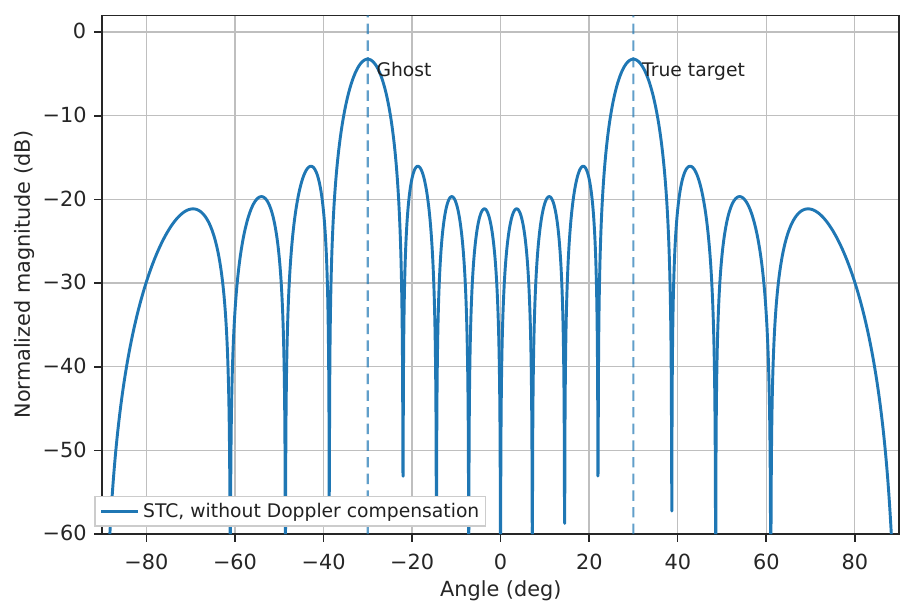}
\caption{Ambiguity in the DOA spectrum due to ghost peak for $L=4$, $N_{\mathrm{RX}}=8$, and a moving target with radial velocity $v_1=34.82$~m/s. The target is located at $\hat{\theta}_{1} =30^{\circ}$, while the ghost peak appears at $\hat{\theta}_{\mathrm{ghost},1} =-30^{\circ}$.}
\label{fig:DOA_ghost_30}
\end{figure}

Now, we include Doppler issues related to the proposed STC scheme in the case of a non-stationary target. For the following Doppler-effect analysis and compensation, we retain the general setting $L=2K$ and the indices $l\in\mathcal L_{\mathrm{ret}}$. The decoded repetitions are accumulated during range correlation as in \eqref{eq:range_proc_STC}. For a moving target, the repetition-dependent Doppler phases in $\mathbf D_K(f_{D,1})$ can make this accumulation destructive. As discussed in Section~\ref{sec:range_sidelobe_analysis}, the Doppler frequency shift leads to non-ideal auto- and cross-correlation, which increases range sidelobes, thereby reducing the radar's dynamic range capability. Furthermore, we will also show that the Doppler effect causes a ghost target to appear in the proposed STC-based processing scheme because the orthogonality of the designed space-time codes is disturbed. Adapting \eqref{eq:STC_received}, \eqref{eq:STC_received_rewrite} and \eqref{eq:STC_decoding}, to include the Doppler effect, we obtain 
\begin{equation}
\begin{aligned}
\widetilde{\mathbf y}_j[m,n]
&=\left(\mathbf X^*\otimes\mathbf I_{K-1}\right)
\mathbf y_j^{\mathrm{ret}}[m,n]\\
&=\mathbf M\left(\mathbf c[n-\nu_1]\otimes\mathbf1_{K-1}\right)\\
&\quad\times\tilde\alpha_1\phi_{\mathrm{RX},\theta_1}^{j}
e^{j2\pi f_{D,1}(nT_c+mT_r)} \\
&=\begin{bmatrix}
\widetilde{\mathbf y}_{j,0}[m,n]\\
\widetilde{\mathbf y}_{j,1}[m,n]
\end{bmatrix},
\end{aligned}
\label{eq:STC_received_rewrite_nonideal}
\end{equation}
where 
\begin{align}
\mathbf M
&=\left(\mathbf X^*\otimes\mathbf I_{K-1}\right)
\mathbf D_K(f_{D,1})\left(\mathbf X^T\otimes\mathbf I_{K-1}\right)\notag\\
&=\begin{bmatrix}m_{0,0}&m_{0,1}\\m_{1,0}&m_{1,1}\end{bmatrix}\otimes\operatorname{diag}\!\left(e^{j\delta_D},\ldots,e^{j(K-1)\delta_D}\right),
\end{align}
\begin{equation}
    \begin{split}
        m_{0,0}&=|s_0|^2+|s_1|^2e^{jK\delta_D},\notag\\
        m_{0,1}&=s_0^*s_1\left(1-e^{jK\delta_D}\right),\notag\\
        m_{1,0}&=s_0s_1^*\left(1-e^{jK\delta_D}\right),\notag\\
        m_{1,1}&=|s_1|^2+|s_0|^2e^{jK\delta_D},\notag
    \end{split}
\end{equation}
where $\delta_D=2\pi f_{D,1}T_d$. To simplify the analysis of the range, velocity, and DOA processing for the proposed STC-based scheme in the presence of Doppler effect, we make the following assumptions:
\begin{equation}
    s_{0} = s_{1}=s \ \text{and} \ c_{0}[n] = c_{1}[n] = c[n].
\end{equation}
As a result, $\mathbf{M}$ becomes 
\begin{equation}
\begin{aligned}
\mathbf M
&=\begin{bmatrix}m_d&m_o\\m_o&m_d\end{bmatrix}\otimes\operatorname{diag}\!\left(e^{j\delta_D},\ldots,e^{j(K-1)\delta_D}\right),
\end{aligned}
\label{eq:M_matrix_with_same_s}
\end{equation}
where 
\[
\begin{aligned}
m_d&=|s|^2\left(1+e^{jK\delta_D}\right),\\
m_o&=|s|^2\left(1-e^{jK\delta_D}\right).
\end{aligned}
\]
We split the analysis of the decoded signal into $\mathrm{TX}_{0}$ and $\mathrm{TX}_{1}$. For the decoded signal from $\mathrm{TX}_{0}$, range processing leads to
\begin{equation}
\begin{aligned}
r_{\mathrm{STC},j,0}[m,\nu]
&=\sum_{b=1}^{K-1}\sum_{n=0}^{N-1}\tilde y_j[m,0,b,n]c[n-\nu]\\
&=\tilde\alpha_1\phi_{\mathrm{RX},\theta_1}^{j}\tilde r[\nu]
\left(m_d+m_o\phi_{\mathrm{TX},\theta_1}\right)\\
&\quad\times e^{j2\pi f_{D,1}mT_r}
\sum_{b=1}^{K-1}e^{jb\delta_D},
\end{aligned}
\label{eq:range_proc_STC_nonideal_0}
\end{equation}
where $\tilde{r}[\nu]=\sum_{n=0}^{N-1} c[n-\nu_{1}]e^{j2\pi f_{D,1}nT_{c}} c[n-\nu]$. Similarly, for the decoded signal from $\mathrm{TX}_{1}$, we have
\begin{equation}
\begin{aligned}
r_{\mathrm{STC},j,1}[m,\nu]
&=\sum_{b=1}^{K-1}\sum_{n=0}^{N-1}\tilde y_j[m,1,b,n]c[n-\nu]\\
&=\tilde\alpha_1\phi_{\mathrm{RX},\theta_1}^{j}\tilde r[\nu]
\left(m_o+m_d\phi_{\mathrm{TX},\theta_1}\right)\\
&\quad\times e^{j2\pi f_{D,1}mT_r}
\sum_{b=1}^{K-1}e^{jb\delta_D},
\end{aligned}
\label{eq:range_proc_STC_nonideal_1}
\end{equation}
Subsequently, velocity processing results in
\begin{equation}
\begin{aligned}
\mathcal R_{\mathrm{STC},j,0}[k,\nu]
&=\sum_{m=0}^{M-1}r_{\mathrm{STC},j,0}[m,\nu]e^{-j\frac{2\pi}{M}mk}\\
&=\tilde\alpha_1\phi_{\mathrm{RX},\theta_1}^{j}\tilde r[\nu]
\left(m_d+m_o\phi_{\mathrm{TX},\theta_1}\right)\\
&\quad\times\sum_{b=1}^{K-1}e^{jb\delta_D}\sum_{m=0}^{M-1}e^{j2\pi\frac mM(Mf_{D,1}T_r-k)},
\end{aligned}
\label{eq:velocity_proc_STC_nonideal_0}
\end{equation}
and
\begin{equation}
\begin{aligned}
\mathcal R_{\mathrm{STC},j,1}[k,\nu]
&=\sum_{m=0}^{M-1}r_{\mathrm{STC},j,1}[m,\nu]e^{-j\frac{2\pi}{M}mk}\\
&=\tilde\alpha_1\phi_{\mathrm{RX},\theta_1}^{j}\tilde r[\nu]
\left(m_o+m_d\phi_{\mathrm{TX},\theta_1}\right)\\
&\quad\times\sum_{b=1}^{K-1}e^{jb\delta_D}\sum_{m=0}^{M-1}e^{j2\pi\frac mM(Mf_{D,1}T_r-k)}.
\end{aligned}
\label{eq:velocity_proc_STC_nonideal_1}
\end{equation}
To continue further with the analysis of DOA processing, we consider an on-grid target detected at $k = \hat{k}_1$ and $\nu=\hat{\nu}_1$ and transform the spatial index $i$ and $j$ into the virtual array index $q=2j+i \in \{0, 1,...,Q-1 \}$ with $Q=2N_{\mathrm{RX}}$ as in Section~\ref{sec:standard_processing}, that is 
\begin{align}
    \mathcal{R}_{\mathrm{STC},j,0}&[\hat{k}_1,\hat{\nu}_1] = \tilde{\alpha}_{1}\tilde{r}[\hat{\nu}_1]M\phi^{j}_{\mathrm{RX},\theta_{1}}\left(m_d+m_o\phi_{\mathrm{TX},\theta_{1}}\right)\notag\\&\quad\times\sum_{b=1}^{K-1}e^{jb\delta_D} \notag \\ 
    &=A \left(m_{d}e^{j2j\psi} + m_{o}e^{j(2j+1)\psi} \right) \notag \\
    &=A \left(m_{d}e^{jq\psi} + m_{o}e^{jq\psi}e^{j\psi} \right) \ \text{for even} \ q,
\label{eq:even_q}
\end{align}
and 
\begin{align}
    \mathcal{R}_{\mathrm{STC},j,1}&[\hat{k}_1,\hat{\nu}_1] = \tilde{\alpha}_{1}\tilde{r}[\hat{\nu}_1]M\phi^{j}_{\mathrm{RX},\theta_{1}}\left(m_o+m_d\phi_{\mathrm{TX},\theta_{1}}\right)\notag\\&\quad\times\sum_{b=1}^{K-1}e^{jb\delta_D} \notag \\
    &=A \left(m_{o}e^{j2j\psi} + m_{d}e^{j(2j+1)\psi} \right) \notag \\
    &=A \left(m_{o}e^{jq\psi}e^{-j\psi} + m_{d}e^{jq\psi} \right) \ \text{for odd} \ q,
\label{eq:odd_q}
\end{align}
where $A = \tilde{\alpha}_{1}\tilde{r}[\hat{\nu}_1]M\sum_{b=1}^{K-1}e^{jb\delta_D}$ and $\psi=\pi \sin(\theta_{1})$. These range-velocity-processed samples (for $i=0,1$) can be unified by noting that 
\begin{equation}
\left\{
    \begin{matrix} 
       e^{j\psi}, \ q \ \text{even} \\
       e^{-j\psi}, \ q \ \text{odd} 
    \end{matrix}
    \right.
    = \cos(\psi)+j(-1)^{q}\sin({\psi}), 
\label{eq:fact_1}   
\end{equation}
and
\begin{equation}
       (-1)^{q} = e^{j\pi q}.
\label{eq:fact_2}   
\end{equation}
As a result, by using \eqref{eq:fact_1} and \eqref{eq:fact_2}, we can express \eqref{eq:even_q} and \eqref{eq:odd_q} in terms of the virtual array index $q$ as 
\begin{equation}
\begin{split}
        \mathcal{R}_{\mathrm{STC},q}[\hat{k}_1,\hat{\nu}_1]&=A \bigl(m_{d}e^{jq\psi} + m_{o}e^{jq\psi}(\cos(\psi) \\
        & \hspace{.03\columnwidth} + je^{j\pi q}\sin({\psi})) \bigr) \\
        &= A \bigl((m_{d} + m_{o}\cos(\psi))e^{jq\psi} \\
        & \hspace{.03\columnwidth} +jm_{o}\sin({\psi})e^{jq(\psi+\pi)} \bigr).
\end{split}
\end{equation}
Finally, the DOA spectrum of the target is obtained by
\begin{equation}
    \begin{split}
        & \mathcal{R}_{\mathrm{STC},\omega}[\hat{k}_1,\hat{\nu}_1] = \sum_{q=0}^{Q-1} \mathcal{R}_{\mathrm{STC},q}[\hat{k}_1,\hat{\nu}_1]e^{-j2\pi\frac{q}{Q}\omega} \\ 
       &= \underbrace{A(m_{d} + m_{o}\cos(\psi))\sum_{q=0}^{Q-1}e^{j2\pi\frac{q}{Q}(\frac{Q}{2\pi}\psi-\omega)}}_{\mathrm{true\ target}} \\
       & \hspace{.03\columnwidth} + \underbrace{jAm_{o}\sin({\psi})\sum_{q=0}^{Q-1}e^{j2\pi\frac{q}{Q}(\frac{Q}{2\pi}(\psi+\pi)-\omega)}}_{\mathrm{ghost\ target}}.
    \end{split}
    \label{eq:DOA_proc_ambiguity}
\end{equation}
It can be observed from \eqref{eq:DOA_proc_ambiguity} that the DOA spectrum contains true and ghost spatial components centered at
\begin{equation}
    \begin{split}
        \hat{\theta}_1 &= \arcsin\left(\frac{\psi}{\pi} \right), \\
        \hat{\theta}_{\mathrm{ghost},1} &= \left \{
    \begin{matrix}
    \arcsin \left(\frac{1}{\pi}(\psi-\pi) \right), &  \psi \geq 0, \\[3pt]
    \arcsin \left(\frac{1}{\pi}(\psi+\pi) \right), &   \psi < 0,
    \end{matrix}
    \right.
    \end{split}
    \label{eq:theta_true_and_ghost}
\end{equation}
where a ghost peak shows at $\hat{\theta}_{\mathrm{ghost},1}$. Consequently, a unique DOA estimation of the target would be difficult to achieve, especially when multiple targets are present and the received signal-to-noise ratio (SNR) is poor. Fig.~\ref{fig:DOA_ghost_30} shows the ambiguity issue due to a ghost peak in which the DOA of the true target cannot be determined without any additional information. 

\begin{figure}[!t]
\centering
\includegraphics[width=0.830\columnwidth,trim=0 7.75bp 0 1.75bp,clip]{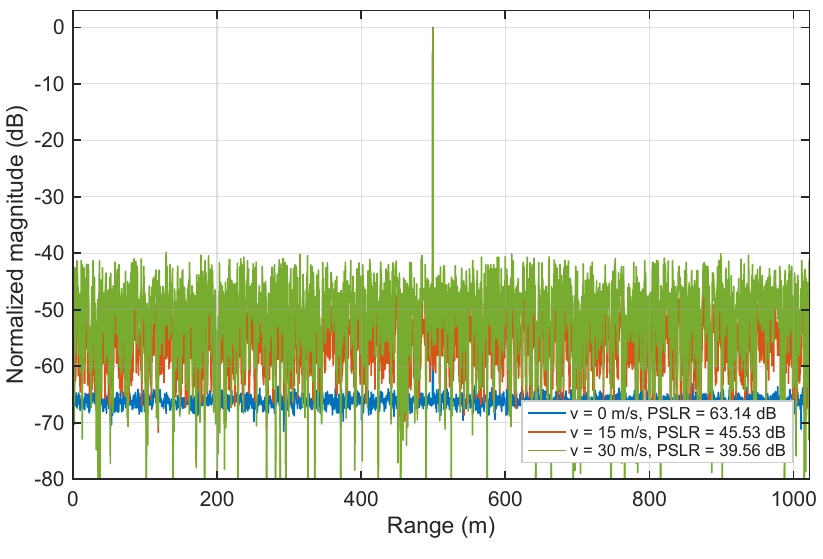}
\caption{Range sidelobes of a target in STC-based processing with different velocities.}
\label{fig:STC_range_sidelobes}
\end{figure}

We can also observe the range sidelobes of the target in the STC-based processing with the Doppler effect. By setting $i=0$ and assuming that the target is located at Doppler bin $k=\hat{k}_1$, its range profile can be expressed by
\begin{equation}
\begin{aligned}
\mathcal R_{\mathrm{STC},j,0}[\hat k_1,\nu]
&=\tilde\alpha_1\phi_{\mathrm{RX},\theta_1}^{j}M\tilde r[\nu]\\
&\quad\times\left(m_d+m_o e^{j\psi}\right)
\sum\nolimits_{b=1}^{K-1}e^{jb\delta_D}.
\end{aligned}
\label{eq:range_sidelobes_STCwithDoppler}
\end{equation}
It can be seen from \eqref{eq:range_sidelobes_STCwithDoppler} that the STC-based processing does not resolve the Doppler shift in the auto-correlation term (i.e., $\tilde{r}[\nu]=\sum_{n=0}^{N-1} c[n-\nu_{1}]e^{j2\pi f_{D,1}nT_{c}} c[n-\nu]$), and it causes high range sidelobes as shown in Fig.~\ref{fig:STC_range_sidelobes}. Furthermore, the addition of $\tilde{r}[\nu]$ in \eqref{eq:range_sidelobes_STCwithDoppler} can be destructive for some DOA and $\delta_{D}$ settings such that $m_{d}+m_{o} e^{j\psi}\approx0$, thus the peak of the target can be masked in low SNR scenarios. For example, let us consider $L=4$ ($K=2$) and a worst-case scenario when $\theta_{1}= 30^{\circ}$ and $\delta_{D}= -\pi/4$. Therefore, 
\begin{equation}
\begin{split}
m_d&=|s|^2\left(1+e^{-j\pi/2}\right)=(1-j)|s|^2,\\
m_o e^{j\psi}&=j|s|^2\left(1-e^{-j\pi/2}\right)\\
&=(-1+j)|s|^2=-m_d,
\end{split}
\end{equation}
and we lose the observable peak of the target in the range profile since $m_{d}+m_{o} e^{j\psi}=0$. In Section~\ref{sec:Doppler_mitigation}, we discuss a technique to combat the issues associated with the Doppler effect in the STC-based processing.

\section{Doppler-Compensated STC-Based Processing}

\label{sec:Doppler_mitigation}
From \eqref{eq:STC_received_rewrite_nonideal}, it is straightforward to see that using $\mathbf X^{*}$ for each retained pair without Doppler compensation is not an optimal decoding strategy since the resulting decoded matrix $\mathbf{M}$ is not a diagonal matrix. Consequently, the proposed STC-based processing performs poorly, which is indicated by the appearance of a ghost target during DOA processing in \eqref{eq:DOA_proc_ambiguity} and range sidelobes in \eqref{eq:range_sidelobes_STCwithDoppler}. 
By observing \eqref{eq:STC_received_rewrite_nonideal}, it is clear that the optimal decoding matrix should also invert the diagonal Doppler matrix, resulting in $\mathbf M=\beta\mathbf I_{L-2}$, and thus eliminating the off-diagonal terms which cause the issues mentioned in Section~\ref{sec:Doppler_issues_STC_proc}.  
However, the main challenge is determining the unknown velocity of the targets. To optimally use the STC-based processing in the presence of Doppler, we propose a technique called Doppler-compensated (DC) STC-based processing, where the velocity processing is performed first, similar to the processing presented in~\cite{Xu2019_DopplerPMCW}. 


Now, we present a modified processing scheme to mitigate the Doppler effect by estimating the velocity information at the beginning of the processing chain. 

\subsubsection{Doppler estimation before range processing}

We first apply an $M$-point DFT along the slow-time index $m$ for every chip $n$ and retained repetition $l\in\mathcal L_{\mathrm{ret}}$,
\begin{equation}
Y_j[k,l,n]
=
\sum_{m=0}^{M-1}
y_j[m,l,n]e^{-j2\pi mk/M}.
\label{eq:doppler_first_dft}
\end{equation}
Substituting \eqref{eq:STC_received} into
\eqref{eq:doppler_first_dft} gives
\begin{align}
Y_j[k,l,n]
&=
\tilde{\alpha}_1
\phi_{\mathrm{RX},\theta_1}^{j}
u_l[n]
e^{j2\pi f_{D,1}(nT_c+lT_d)}
\nonumber\\
&\quad\times
\underbrace{
\sum_{m=0}^{M-1}
e^{j2\pi m(f_{D,1}T_r-k/M)}
}_{G_M(f_{D,1}T_r-k/M)},
\label{eq:doppler_first_output}
\end{align}
where $u_l[n]$ is the weighted sum of the two transmit sequences in the coding interval, and it is split between $l=b$ and $l=K+b$ with $b=1,\ldots,K-1$ as
\[
\begin{aligned}
u_b[n]&=s_0c_0[n-\nu_1]
 +s_1\phi_{\mathrm{TX},\theta_1}c_1[n-\nu_1],\\
u_{K+b}[n]&=-s_1^*c_0[n-\nu_1]
 +s_0^*\phi_{\mathrm{TX},\theta_1}c_1[n-\nu_1].
\end{aligned}
\] The last term in \eqref{eq:doppler_first_output} is a finite geometric sum, which can be expressed as a phase-shifted Dirichlet kernel
\begin{equation}
G_M(\epsilon)
=
e^{j\pi(M-1)\epsilon}
\frac{\sin(\pi M\epsilon)}
{\sin(\pi\epsilon)},
\label{eq:dirichlet_kernel}
\end{equation}
with $G_M(0)=M$. Its magnitude is maximized when $k/M \approx f_{D,1}T_r$. Therefore, a
Doppler-bin estimate can be obtained as
\begin{equation}
\hat{k}_1 =\arg\max_k E[k], \qquad \hat{f}_{D,1} = \frac{\hat{k}_1}{MT_r},
\label{eq:doppler_estimate}
\end{equation}
where we use a simple non-coherent energy statistic, defined as
\begin{equation}
E[k]
=
\sum_{l\in\mathcal L_{\mathrm{ret}}}\sum_{n=0}^{N-1}
\left|Y_j[k,l,n]\right|^2.
\label{eq:doppler_energy}
\end{equation}
Since the Doppler phase progression is identical across the receive antennas and the receive-array response only introduces an antenna-dependent phase factor, the Doppler-bin estimate can be obtained from any single receive antenna. Alternatively, non-coherent integration across multiple receive antennas can be employed to improve robustness in noise. It is important to note that an off-grid target can cause inaccurate Doppler estimation, thus resulting in an imperfect Doppler compensation. To improve the accuracy of Doppler estimation, a slow-time window can be included in \eqref{eq:doppler_first_dft} to reduce
Doppler sidelobes, and a zero-padding can also be used to refine $\hat{f}_{D,1}$ if the target is located between DFT bins. Let us define a Doppler estimation error as 
\begin{equation}
\Delta f_{D,1}=f_{D,1}-\hat{f}_{D,1}, 
\end{equation}
and this error is bounded by
\begin{equation}
|\Delta f_{D,1}|\leq\frac{1}{2MT_r}, 
\label{eq:Doppler_error_bound}
\end{equation}
where we assume that the Doppler estimate is obtained by selecting the DFT bin nearest to the true Doppler frequency, without noise-induced bin errors, and \eqref{eq:Doppler_error_bound} considers the worst-case of half-bin rounding errors.  Zero-padding refines the frequency grid and can reduce Doppler estimation error since it lowers the upper bound on \eqref{eq:Doppler_error_bound}, but it does not improve the Doppler resolution set by the frame duration.

\subsubsection{Joint fast-time and STC-slot Doppler compensation}

For the estimated Doppler frequency $\hat{f}_{D,1}$, and its corresponding Doppler bin $\hat{k}_1$, the compensated version of \eqref{eq:doppler_first_output} is 
\begin{equation}
\overline Y_j[l,n]
=Y_j[\hat k_1,l,n]
e^{-j2\pi\hat f_{D,1}(nT_c+lT_d)}, \quad l\in\mathcal L_{\mathrm{ret}}.
\label{eq:compensated_samples}
\end{equation}
The $nT_c$ term removes the Doppler phase across the PMCW chips, while the $lT_d$ term removes the relative Doppler phase between the STC slots. The latter is essential for restoring the orthogonality of the STC coding matrix. Considering Doppler estimation error, the compensated vector, stacking $\overline Y_j[l,n]$ in increasing order of $l\in\mathcal L_{\mathrm{ret}}$, can be expressed as
\begin{equation}
\begin{aligned}
\overline{\mathbf Y}_j[n]
&=A_{D,1} e^{j2\pi\Delta f_{D,1}nT_c}
\mathbf D_K(\Delta f_{D,1})\\
&\quad\times\left(\mathbf X^T\otimes\mathbf1_{K-1}\right)
\mathbf c[n-\nu_1],
\end{aligned}
\label{eq:residual_doppler_model}
\end{equation}
where
\begin{equation}
A_{D,1}=\tilde{\alpha}_1\phi_{\mathrm{RX},\theta_1}^{j}G_M (f_{D,1}T_r- \hat{k}_1/M ).
\label{eq:AD}
\end{equation}
Thus, an off-grid target can still be compensated effectively, and the remaining distortion is controlled by the Doppler estimation error $\Delta f_{D,1}$.

\subsubsection{Restoration of STC orthogonality}

Applying the same STC decoder as in \eqref{eq:STC_decoding} gives
\begin{equation}
\begin{aligned}
&\widetilde{\mathbf Y}_j[n]
=\left(\mathbf X^*\otimes\mathbf I_{K-1}\right)\overline{\mathbf Y}_j[n]\\
&=A_{D,1} e^{j2\pi\Delta f_{D,1}nT_c}\mathbf M(\Delta f_{D,1})\left(\mathbf c[n-\nu_1]\otimes\mathbf1_{K-1}\right),
\end{aligned}
\label{eq:compensated_stc_decode}
\end{equation}
where
\begin{equation}
\begin{aligned}
\mathbf M(\Delta f_{D,1})
&=\left(\mathbf X^*\otimes\mathbf I_{K-1}\right)\mathbf D_K(\Delta f_{D,1})\left(\mathbf X^T\otimes\mathbf I_{K-1}\right).
\end{aligned}
\label{eq:residual_M}
\end{equation}
For a perfect Doppler estimation, i.e., $\Delta f_{D,1}=0$, we have $\mathbf D_K(0)=\mathbf I_{L-2}$ and $\mathbf M(0)=\beta\mathbf I_{L-2}$, with $\beta=|s_0|^2+|s_1|^2$, which shows that the STC orthogonality is restored for every retained repetition.

\subsubsection{Range processing after Doppler compensation}
For simplicity, we again assume $c_0[n]=c_1[n]=c[n]$ and a perfect Doppler compensation. The two
decoded channels become
\begin{align}
\widetilde{\mathbf Y}_j[n]
&=\begin{bmatrix}\widetilde{\mathbf Y}_{j,0}[n]\\\widetilde{\mathbf Y}_{j,1}[n]\end{bmatrix}=A_{D,1}\beta\begin{bmatrix}c[n-\nu_1]\mathbf1_{K-1}\\
\phi_{\mathrm{TX},\theta_1}c[n-\nu_1]\mathbf1_{K-1}\end{bmatrix}.
\label{eq:decoded_tx1}
\end{align}
Here, $\widetilde{\mathbf Y}_{j,i}[n]$ contains the samples $\widetilde Y_{j,i}[b,n]$ for $b=1,\ldots,K-1$. Therefore, the range correlation with repetition accumulation for the $i$-th decoded transmit channel is 
\begin{equation}
\begin{split}
    r_{\mathrm{DC},j,i}[\nu] &= \sum_{b=1}^{K-1}\sum_{n=0}^{N-1}\widetilde Y_{j,i}[b,n]c[n-\nu] \\
    &= (K-1)A_{D,1}\beta \phi_{\mathrm{TX},\theta_1}^{i}\tilde{r}[\nu],
\end{split}
\label{eq:range_after_compensation}
\end{equation}
where
\begin{equation}
\tilde{r}[\nu]=\sum_{n=0}^{N-1}c[n-\nu_1]c[n-\nu] \notag
\end{equation}
is the Doppler-free PACF, while \eqref{eq:range_proc_STC_nonideal_0} and \eqref{eq:range_proc_STC_nonideal_1} contain $e^{j2\pi f_{D,1}nT_c}$ in their range correlations. For an $m$-sequence, the PACF in \eqref{eq:mseq_PACF} is consequently recovered.

\subsubsection{DOA processing}
At the detected range bin $\nu=\hat{\nu}_{1}$, we can arrange the samples defined in \eqref{eq:range_after_compensation} $\forall i,j$ into virtual array index $q=2j+i$ as 
\begin{equation}
\mathcal{R}_{\mathrm{DC},q}[\hat{\nu}_{1}]=A \tilde{r}[\hat{\nu}_{1}]e^{jq\psi},
\qquad
q=0,\ldots,Q-1,
\label{eq:compensated_virtual_array}
\end{equation}
where $A = \tilde{\alpha}_1 G_M(f_{D,1}T_r - \hat{k}_1/M)(K-1)\beta$, $\psi=\pi\sin(\theta_1)$, and $Q=2N_{\mathrm{RX}}$. Then, the corresponding spatial DFT is
\begin{equation}
\mathcal{R}_{\mathrm{DC},\omega} =A\tilde r[\hat{\nu}_1] \sum_{q=0}^{Q-1} e^{j\frac{2\pi q}{Q} \left(\frac{Q}{2\pi}\psi-\omega\right)}.
\label{eq:compensated_spatial_dft}
\end{equation}
Unlike \eqref{eq:DOA_proc_ambiguity}, \eqref{eq:compensated_spatial_dft} does not contain the second spatial sum centered at $\psi+\pi$, because $m_o=0$. Consequently, the ghost peak is suppressed and only the true target peak remains located at $\hat{\theta}_1$ given in \eqref{eq:theta_true_and_ghost}. 

For the case where $\Delta f_{D,1}\neq0$ and assuming $s_0=s_1$, we can define the magnitude ratio between the undesired and desired STC coefficients as
\begin{equation}
\begin{aligned}
\left|\frac{m_o(\Delta f_{D,1})}{m_d(\Delta f_{D,1})}\right|
&=\left|\frac{1-e^{j2\pi K\Delta f_{D,1}T_d}}
{1+e^{j2\pi K\Delta f_{D,1}T_d}}\right|\\
&=\left|\tan(\pi K\Delta f_{D,1}T_d)\right|,
\end{aligned}
\label{eq:leakage_ratio}
\end{equation}
where \eqref{eq:leakage_ratio} quantifies the relationship between Doppler estimation error and the off-diagonal residual STC leakage, such that for $\pi K|\Delta f_{D,1}|T_d\ll1$, 
\begin{equation}
\left|\frac{m_o(\Delta f_{D,1})}{m_d(\Delta f_{D,1})}\right|
\approx
\pi K|\Delta f_{D,1}|T_d,
\label{eq:off_diagonal_error_ratio}
\end{equation}
showing that the improved accuracy of Doppler-frequency estimation directly reduces the residual STC leakage and the associated ghost component.

\subsubsection{Multiple-target implementation}
\begin{figure}[!t]
\centering
\includegraphics[width=1.000\columnwidth,trim=0 4.75bp 0 8.50bp,clip]{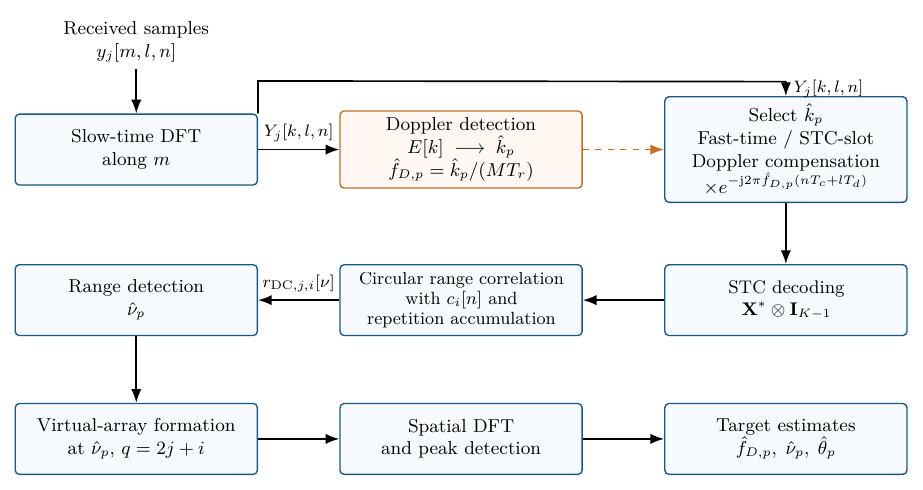}
\caption{Block diagram of the proposed Doppler-compensated STC-based processing. Doppler estimation precedes joint fast-time and STC-slot compensation, followed by STC decoding, range correlation with repetition accumulation, and DOA estimation. Processing is performed independently for each detected Doppler bin.}
\label{fig:Doppler_compensated_STC}
\end{figure}

For $P>1$, targets generally have different Doppler frequencies, and therefore a single-frequency Doppler compensation cannot simultaneously remove the fast-time and coding slot phase rotations of all targets. To make this explicit, the retained sample in the multiple-target case can be written as
\begin{equation}
\begin{aligned}
\mathbf y_j^{\mathrm{ret}}[m,n]
&=\sum_{p=1}^{P}\tilde\alpha_p\phi_{\mathrm{RX},\theta_p}^{j}
e^{j2\pi f_{D,p}(nT_c+mT_r)}\\
&\quad\times\mathbf D_K(f_{D,p})
\left(\mathbf X^T\otimes\mathbf1_{K-1}\right)
\mathbf c_p[n-\nu_p],
\end{aligned}
\label{eq:multi_rx_vector}
\end{equation}
where
\begin{equation}
\mathbf c_p[n-\nu_p]=
 \begin{bmatrix}
 c_0[n-\nu_p]\\
 \phi_{\mathrm{TX},\theta_p}c_1[n-\nu_p]
 \end{bmatrix}.
 \label{eq:multi_cp}
\end{equation}
Applying the $M$-point DFT along the slow-time index $m$ and stacking $Y_j[k,l,n]$ in increasing order of $l\in\mathcal L_{\mathrm{ret}}$ gives
\begin{equation}
\begin{aligned}
\mathbf Y_j[k,n]
&=\sum_{p=1}^{P}\tilde\alpha_p\phi_{\mathrm{RX},\theta_p}^{j}
G_M(\epsilon_{p,k})e^{j2\pi f_{D,p}nT_c}\\
&\quad\times\mathbf D_K(f_{D,p})
\left(\mathbf X^T\otimes\mathbf1_{K-1}\right)
\mathbf c_p[n-\nu_p],
\end{aligned}
\label{eq:multi_doppler_dft}
\end{equation}
where
\begin{equation}
 \epsilon_{p,k}= f_{D,p}T_r-\frac{k}{M}.
 \label{eq:multi_eps}
\end{equation}
Then, we use the metric $E[k]$ as in   \eqref{eq:doppler_energy}, and CFAR or other adaptive detection techniques can be used to detect the $P$ peak locations. For the $p$-th detected bin $\hat k_p$, its Doppler frequency is estimated as
\begin{equation}
 \hat f_{D,p}=\frac{\hat k_p}{MT_r},
 \label{eq:multi_fd_est}
\end{equation}
and the bin-dependent compensation is performed as
\begin{equation}
 \overline{\mathbf{Y}}_{j}^{(p)}[n]
 =e^{-j2\pi\hat f_{D,p}nT_c}
 \mathbf D_K^{-1}(\hat f_{D,p})\mathbf{Y}_j[\hat k_p,n],
 \label{eq:multi_comp}
\end{equation}
and we consider that $G_M(\epsilon_{p,k})=0$ for $k\neq\hat k_p$ in \eqref{eq:multi_comp} by assuming that targets lie exactly on distinct DFT bins, and they are sufficiently separated in the Doppler spectrum. Therefore, we have distinct compensated samples $\overline{\mathbf{Y}}_{j}^{(1)},\ldots,\overline{\mathbf{Y}}_{j}^{(P)}$, and they are subsequently STC-decoded, range-correlated with repetition accumulation, and spatially processed independently. These $P$ branches can be evaluated sequentially or in parallel. For targets on distinct DFT bins, substituting~\eqref{eq:multi_doppler_dft} into~\eqref{eq:multi_comp} yields
\begin{equation}
\begin{aligned}
\overline{\mathbf Y}_j^{(p)}[n]
&=A_{D,p} e^{j2\pi\Delta f_{D,p}nT_c}
\mathbf D_K(\Delta f_{D,p})\\
&\quad\times\left(\mathbf X^T\otimes\mathbf1_{K-1}\right)
\mathbf c_p[n-\nu_p],
\end{aligned}
\label{eq:multi_comp_expanded}
\end{equation}
where
\begin{equation}
 A_{D,p}= \tilde\alpha_p\phi_{\mathrm{RX},\theta_p}^{j}
 G_M\!\left(f_{D,p}T_r-\frac{\hat k_p}{M}\right),
 \label{eq:multi_Apr}
\end{equation}
and
\begin{equation}
 \Delta f_{D,p}= f_{D,p}-\hat f_{D,p}.
 \label{eq:multi_residual}
\end{equation}
Applying the STC decoder gives
\begin{align}
&\widetilde{\mathbf Y}_j^{(p)}[n]
=\left(\mathbf X^*\otimes\mathbf I_{K-1}\right)\overline{\mathbf Y}_j^{(p)}[n]\notag\\
&=A_{D,p} e^{j2\pi\Delta f_{D,p}nT_c}\mathbf M(\Delta f_{D,p})\left(\mathbf c_p[n-\nu_p]\otimes\mathbf1_{K-1}\right),
\label{eq:multi_stc_decode}
\end{align}
where
\begin{equation}
\begin{aligned}
\mathbf M(\Delta f_{D,p})
&=\left(\mathbf X^*\otimes\mathbf I_{K-1}\right)\mathbf D_K(\Delta f_{D,p})\left(\mathbf X^T\otimes\mathbf I_{K-1}\right),
\end{aligned}
\label{eq:multi_M}
\end{equation}
and $\mathbf M(\Delta f_{D,p})=\beta\mathbf I_{L-2}$ if $\Delta f_{D,p}=0$. The range and DOA processing are then performed on \eqref{eq:multi_stc_decode}, similar to \eqref{eq:range_after_compensation}--\eqref{eq:compensated_spatial_dft}. Fig. \ref{fig:Doppler_compensated_STC} shows the complete processing chain of the proposed Doppler-compensated STC-based processing for MIMO PMCW radar.

The proposed framework can be extended to $N_{\mathrm{TX}}>2$. For radar sensing with known transmitted signals, any full-rank coding matrix $\mathbf X\in\mathbb C^{N_{\mathrm{TX}}\times K_s}$ such that $\mathbf X_{\mathrm{dec}}\mathbf X^{T}=\mathbf I$ can separate the transmitted sequences, with $\mathbf{X}_{\mathrm{dec}}=(\mathbf{X}^{*}\mathbf{X}^{T})^{-1}\mathbf{X}^{*}$ used for decoding with $K_s \geq N_{\mathrm{TX}}$. In the case of integrated sensing and communications (ISAC) applications, unknown information symbols can be transmitted in $\mathbf{X}_m$ for every subframe $m$, thus an appropriate space-time block code and symbol decoder are also required since the symbols needs to be estimated at the communication receiver. The orthogonal matrix designs for generalized space-time coding are presented in \cite{Moon2005_ECC}. We will consider the ISAC scenario as a future work.

\section{Simulation Results}
\label{sec:simulation}
Numerical simulations are conducted to validate the main properties of the proposed space-time-coding (STC)-based MIMO PMCW processing with/without Doppler compensation technique. The Alamouti symbols are $s_{0,m}=s_{1,m}=1$ for all subframes, giving $\beta=2$. The experiments are organized to isolate three effects: the range sidelobes caused by non-zero PCCF, the degradation caused by the Doppler phase progression across PMCW chips and STC slots, and the performance of the complete Doppler-compensated STC processing in a multiple-target scenario. To provide comparison of the range sidelobe performance, the existing techniques such as conventional CDM-MIMO and code diversity \cite{Bauduin2021_CodeDiversity} are used. It is important to note that the STC-based processing discards the first repetition of each coding interval, while the code diversity scheme discards one repetition per subframe \cite{Bauduin2021_CodeDiversity}. Thus, for the same subframe duration, conventional CDM-MIMO uses $L$ repetitions, STC uses $L-N_{\mathrm{TX}}$, and code diversity uses $L-1$. For the simulated $L=4$ and $N_{\mathrm{TX}}=2$, these correspond to four, two, and three repetitions, respectively. Despite ignoring the additive noise in our received signal models for analysis simplicity, we incorporate complex Gaussian noise with variance $\sigma^2=0.0025$ into the received signal samples to produce more realistic results. These noise samples
are independent across chips, repetitions, and receive antennas. As a performance metric for the range sidelobes, we define the peak sidelobe ratio (PSLR) as
\begin{equation}
    \mathrm{PSLR}
    =20\log_{10}\!\left(\frac{|r[\nu_{p}]|}{\max_{\nu\neq\nu_{p}}|r[\nu]|}\right).
\end{equation}

For a length-$N$ $m$-sequence, the periodic autocorrelation function (PACF) consists of a main peak of $N$ and sidelobes of magnitude one. Consequently, in the absence of Doppler distortion, its PSLR is
\begin{equation}
    \mathrm{PSLR}_{m\text{-seq}}
    =20\log_{10}\!\left(\frac{1}{1/N}\right)
    =66.22~\text{dB}. 
    \label{eq:m_seq_PSLR}
\end{equation}
The range, Doppler, and DOA spectrum is normalized with respect to the target peak. For DOA estimation, the $Q$ virtual-channel samples are zero-padded to $16384$ points. This refines the spatial-frequency grid without changing the angular resolution, and the angular spectra are displayed at $0.05^{\circ}$ intervals. The main simulation parameters are summarized in Table~\ref{tab:params}. At the maximum unambiguous velocity of $34.82~\mathrm{m/s}$, the
maximum range migration over one frame is $|v_{\max}|T_f
\approx 0.0835~\mathrm{m}$, or $0.167$ of a range bin, thus satisfying the fixed-delay approximation in~\eqref{eq:s_down_approx}.

\begin{table}[!t]
\caption{Main Simulation Parameters}
\label{tab:params}
\centering
\footnotesize
\begin{tabular}{ |c|c| }
\hline
Parameter  & Value \\
\hline \hline
Carrier frequency $f_c$ & 79 GHz \\ \hline
Chip duration $T_c$ & 3.33 ns \\ \hline
Code length $N$ & 2047 \\ \hline
Code duration $T_d$ & 6.8165 $\mu$s \\ \hline
Range resolution $\Delta R$ & 0.4995 m \\ \hline
Transmit antennas $N_{\mathrm{TX}}$ & 2 \\ \hline
Receive antennas $N_{\mathrm{RX}}$ & 8 \\ \hline
Virtual channels $Q$ & 16 \\ \hline
Slow-time samples $M$ & 88 \\ \hline
Slow-time interval $T_r$ & $4T_d$ \\ \hline
\end{tabular}
\end{table}

\subsection{Suppression of Cross-Correlation-Induced Range Sidelobes}

The first experiment isolates the range sidelobes caused by non-zero PCCFs. A stationary broadside target is positioned at range $R=499.5$~m. Since $f_D=0$, the Doppler-dependent impairments derived in Section \ref{sec:range_sidelobe_analysis} are absent, and the difference between conventional CDM-MIMO, code diversity, and STC framework is determined only by the MIMO code-separation mechanism.

\begin{figure}[!t]
\centering
\includegraphics[width=0.830\columnwidth,trim=0 7.75bp 0 8.50bp,clip]{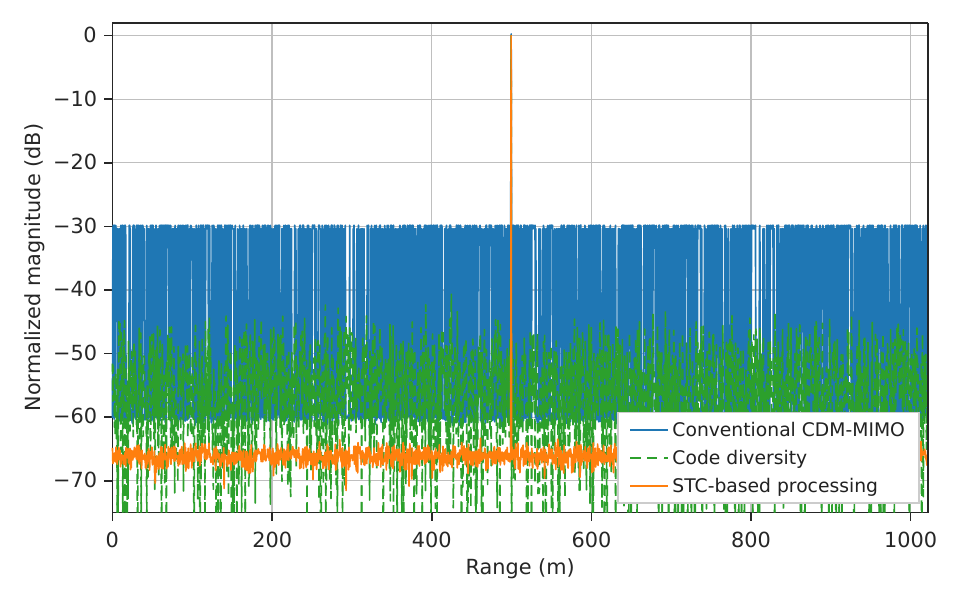}
\caption{Normalized range profiles of a stationary
broadside target at $499.5$~m for conventional CDM-MIMO, code diversity,
and STC-based processing.}
\label{fig:cdm_stc}
\end{figure}

Fig.~\ref{fig:cdm_stc} compares the resulting range cuts. In the conventional CDM-MIMO scheme, correlation with the code assigned to the first transmit antenna contains its PACF together with the PCCF contribution from the second transmit antenna. Its measured PSLR is
$30.06$~dB for the used code sequences since the cross-correlation terms dominate the sidelobe floor even though each individual sequence has an excellent PACF. The code diversity scheme changes the code over slow time, and it reduces the coherent PCCF accumulation, yielding a PSLR of $40.65$~dB. However, the code diversity performance depends on the selected code and the number of subframes. The number of distinct degree-$r$ primitive binary polynomials, each generating an $m$-sequence up to cyclic shift, is
\begin{equation}
    C_N = \frac{\varphi(2^{r}-1)}{r}, \notag
\end{equation}
where $r=\log_{2}(N+1)$, and $\varphi(\cdot)$ is Euler's totient function \cite{Moon2005_ECC}. For the parameters used in this simulation, we have
\begin{equation}
    C_{2047} = \frac{\varphi(2^{11}-1)}{11} = 176. \notag
\end{equation}
Consequently, increasing $M$ or $N_{\mathrm{TX}}$ in this experiment requires another code family if every antenna and subframe must use a different sequence.

In the STC-based processing, the decoding process separates the transmitted components before range correlation, thus the PCCF term is eliminated, and the resulting range profile is limited only by the PACF of the $m$-sequence. Therefore, the theoretical PSLR of $m$-sequence in \eqref{eq:m_seq_PSLR} can be achieved. However, we obtain a PSLR of $63.14$~dB in this experiment due to the addition of Gaussian noise.

\subsection{Doppler Effect and Joint Compensation}
\begin{figure}[!t]
\centering
\includegraphics[width=0.830\columnwidth,trim=0 7.75bp 0 8.50bp,clip]{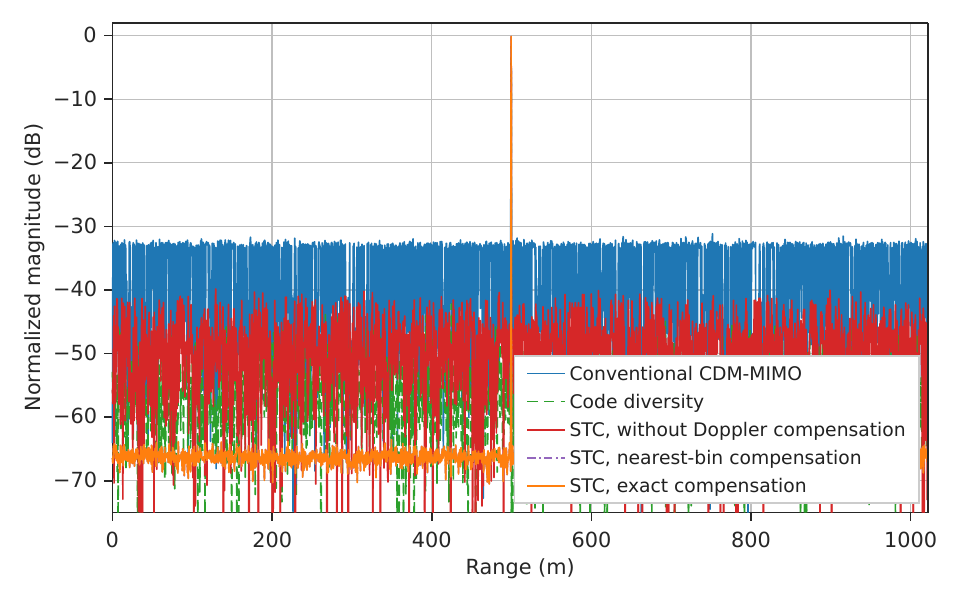}
\caption{Normalized range profile for a target moving at $30$~m/s. The uncompensated Doppler phase substantially increases the range sidelobes. Joint fast-time and STC-slot Doppler compensation restores the PACF-limited range response.}
\label{fig:doppler_range}
\end{figure}

The second experiment considers a broadside target at $R=499.5$~m
moving with radial velocity $v=30$~m/s. The corresponding Doppler frequency introduces the chip-wise phase progression $e^{j2\pi f_D nT_c}$, which distorts the PACF of the code sequence. In the STC framework, the Doppler effect additionally produces a relative phase rotation between the two STC slots, which breaks the orthogonality of the STC coding matrix. Fig.~\ref{fig:doppler_range} compares the range profile of
conventional CDM-MIMO, code diversity, uncompensated STC, and STC
with nearest-bin or exact Doppler compensation. The conventional CDM-MIMO contains both PACF and PCCF contributions and gives a measured PSLR of $28.46$~dB. The code diversity scheme reduces coherent accumulation of code-dependent sidelobes
over slow time, yielding PSLR of $40.52$~dB. The uncompensated STC-based processing gives PSLR of $39.56$~dB, which shows that STC separation alone does not restore the Doppler-free PACF in this moving-target scenario. 

\begin{figure}[!t]
\centering
\includegraphics[width=0.830\columnwidth,trim=0 7.75bp 0 8.50bp,clip]{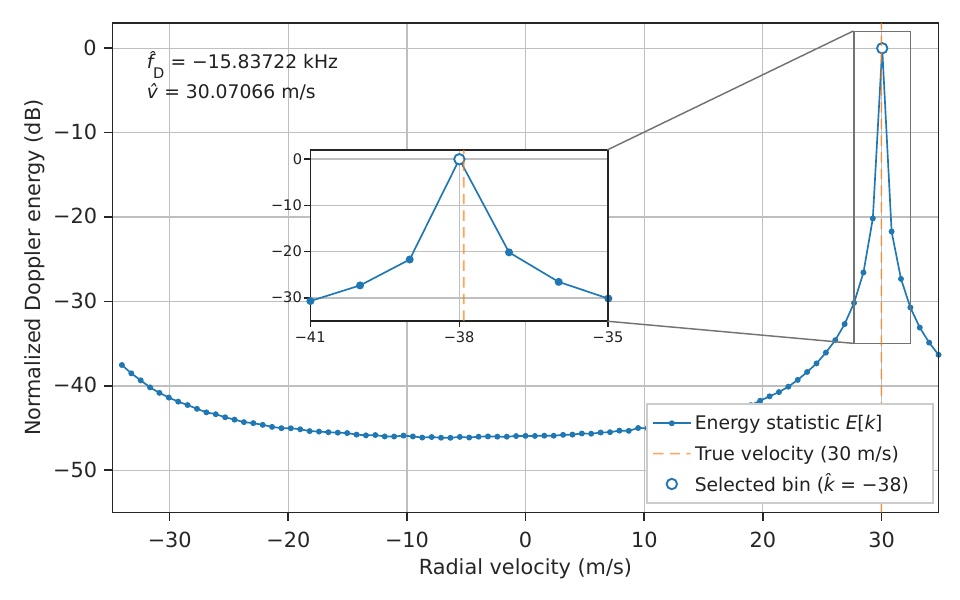}
\caption{Normalized Doppler energy statistic $E[k]$ defined in \eqref{eq:doppler_energy}. The peak selects $\hat{k}=-38$, corresponding to $\hat{f}_D=-15.83722$~kHz and $\hat{v}=30.07066$~m/s.}
\label{fig:doppler_spectrum}
\end{figure}

For the compensated STC framework, we perform the slow-time Doppler processing, prior to STC decoding and range correlation as described in 
Section \ref{sec:Doppler_mitigation}. Fig. \ref{fig:doppler_spectrum} shows the Doppler spectrum by using the energy statistic defined in \eqref{eq:doppler_energy}.
The selected slow-time DFT bin corresponds to
$\hat v=30.07066$~m/s, and it differs slightly from the true velocity due to off-grid target. Nearest-bin compensation uses its frequency $\hat f_D=-2f_c\hat v/c$, which leaves a Doppler estimation error $|\Delta f_{D}|$ of approximately $37.22$~Hz. Exact compensation uses the true target frequency $f_D$ for the chip-wise and STC-slot corrections, while retaining the same selected slow-time DFT bin. With the receiver noise added, the nearest-bin and exact compensation give PSLRs of $63.13$~dB and $63.14$~dB, respectively, and their range cuts nearly coincide.

\begin{figure}[!t]
\centering
\includegraphics[width=0.830\columnwidth,trim=0 8.50bp 0 10.75bp,clip]{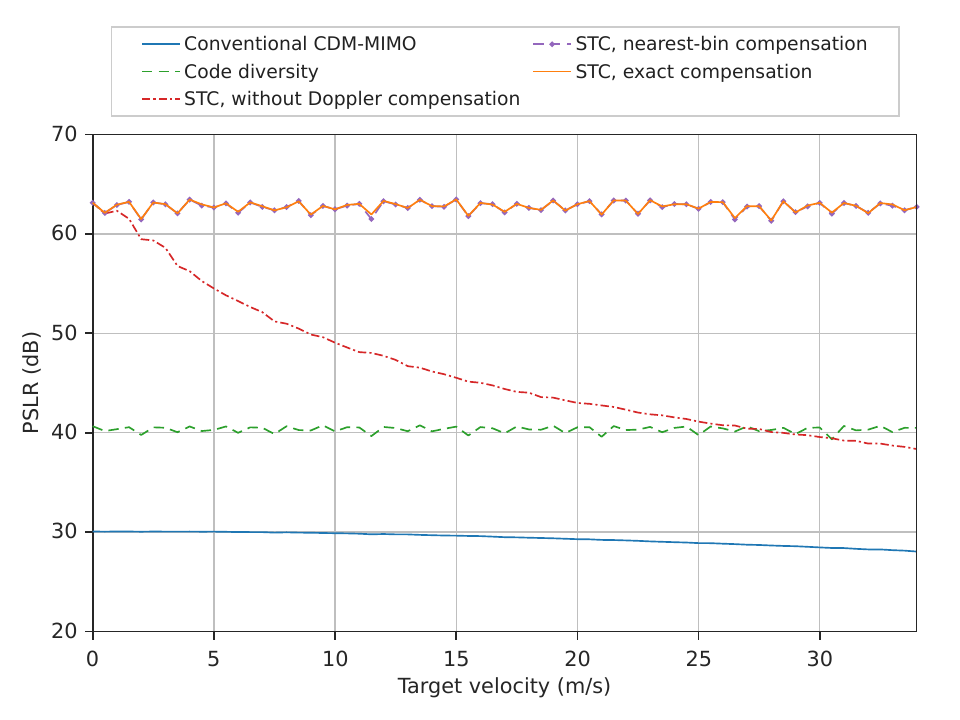}
\caption{PSLR value versus target velocity.}
\label{fig:PSLR_velocity}
\end{figure}

Fig. \ref{fig:PSLR_velocity} compares the range PSLR of conventional CDM-MIMO, code diversity, and STC over target velocities from 0 to 34 m/s. Conventional CDM-MIMO remains cross-correlation limited, while code diversity maintains approximately 40 dB PSLR. Uncompensated STC scheme degrades with velocity, reaching 39.56 dB at 30 m/s because of chip-wise Doppler distortion and loss of decoding orthogonality. Joint Doppler compensation maintains approximately 61.3--63.5~dB PSLR, with nearest-bin compensation closely matching the exact compensation. 

\subsection{Suppression of Doppler-Induced DOA Ghosts}

\begin{figure}[!t]
\centering
\includegraphics[width=0.830\columnwidth,trim=0 7.75bp 0 13.00bp,clip]{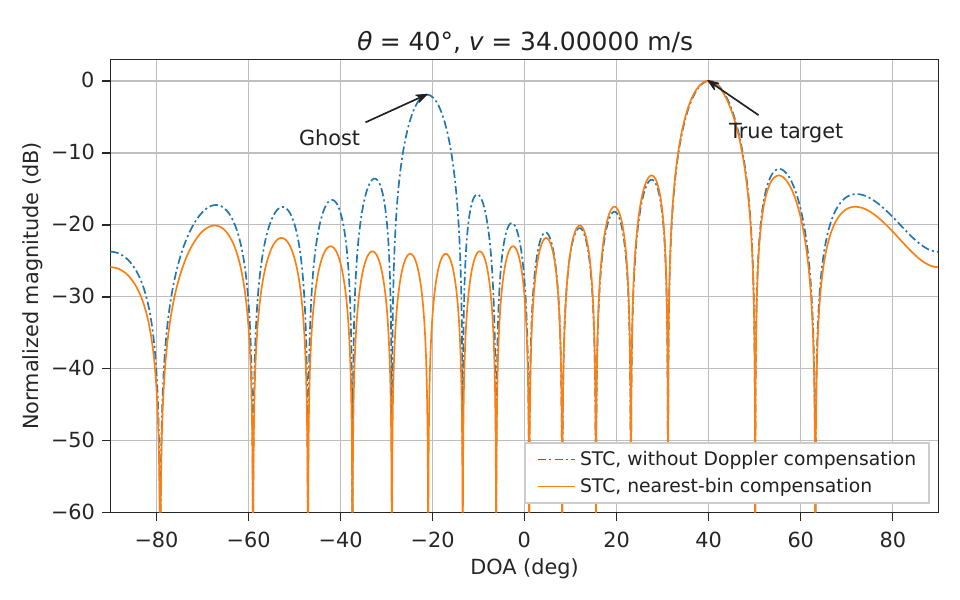}
\caption{Normalized DOA spectra for a target at $\theta=40^{\circ}$ moving at $v=34$~m/s. The blue dashed line indicates a ghost peak at around $-20.93^{\circ}$, while the Doppler-compensated STC framework in the orange solid line suppresses the extra peak.}
\label{fig:doa_ghost_40}
\end{figure}

The third experiment shows the effect of Doppler-induced STC leakage on the DOA estimation. The target 
is located at $\theta=40^{\circ}$ and moves with $v=34$~m/s. From the analysis in Section \ref{sec:Doppler_issues_STC_proc}, the uncompensated STC decoder will produce a ghost peak, in addition to the true target's peak. According to \eqref{eq:theta_true_and_ghost}, the ghost peak appears at $\theta_{\mathrm{ghost}}=-20.93^{\circ}$, and it is verified by the simulation, shown as the blue dashed line in Fig. \ref{fig:doa_ghost_40}. Since the ghost component is only $1.92$~dB below the actual target's peak, we can have an ambiguous DOA estimation.

The STC-based processing with nearest-bin compensation uses the Doppler bin $-43$, corresponding to $\hat v=34.0273$~m/s. Its residual Doppler error is approximately $14.39$~Hz, thus the compensation nearly restores the decoding orthogonality. The orange solid line in Fig. \ref{fig:doa_ghost_40} shows the suppressed ghost component, and only the true target peak remains. The remaining angular sidelobes are primarily due to the finite virtual-array aperture. 

\subsection{Multiple-Target Evaluation}

\begin{table}[!t]
\caption{True and estimated parameters of the multiple-target scenario after Doppler refinement.}
\label{tab:three_targets_true_estimated}
\centering
\footnotesize
\begin{tabular}{|c|c|c|c|c|}
\hline
Target & Type & Range & Velocity & DOA \\
 & & (m) & (m/s) & (deg) \\
\hline
\hline
\multirow{2}{*}{1}
 & True & 149.85 & $-16.22$ & $-30$ \\ \cline{2-5}
 & Estimated & 149.85 & $-16.22$ & $-30$ \\ \hline
\multirow{2}{*}{2}
 & True & 499.50 & $8.31$ & $20$ \\ \cline{2-5}
 & Estimated & 499.50 & $8.31$ & $20$ \\ \hline
\multirow{2}{*}{3}
 & True & 749.25 & $24.14$ & $60$ \\ \cline{2-5}
 & Estimated & 749.25 & $24.14$ & $60.01$ \\ \hline
\end{tabular}
\end{table}

\begin{figure}[!t]
\centering
\includegraphics[width=0.830\columnwidth,trim=0 0.25bp 0 0.25bp,clip]{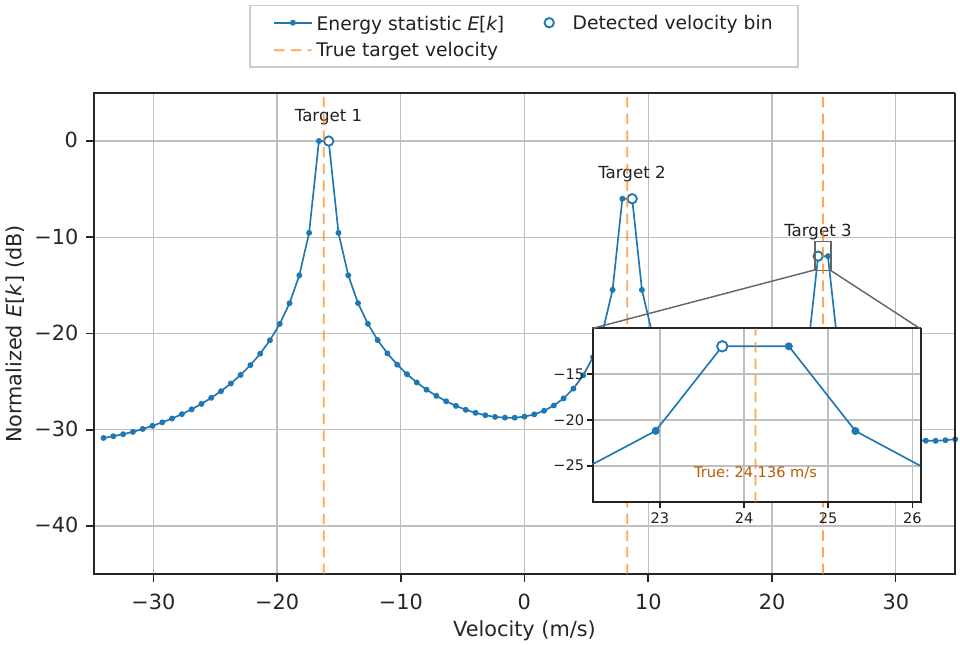}
 \caption{Normalized Doppler energy statistic from the 88-point DFT without windowing or zero-padding. The targets lie halfway between adjacent bins. The dashed lines mark the true velocities and the hollow circles denote the selected Doppler bins.}
\label{fig:Doppler_spectrum_multitarget_unrefined}
\end{figure}

Finally, the complete processing chain is evaluated in a multiple-target scenario. Three targets with distinct ranges, velocities, angles, and amplitudes are simulated. Their true and estimated parameters are listed in Table~\ref{tab:three_targets_true_estimated}. The slow-time DFT is first applied to the received space-time-coded samples. Fig. \ref{fig:Doppler_spectrum_multitarget_unrefined} shows the Doppler spectrum without windowing or zero-padding. For this noisy realization, the selected signed Doppler bins are $20$, $-11$, and $-30$, corresponding to velocities of $-15.8266$, $8.7046$, and $23.7399$~m/s, respectively.

\begin{figure}[!t]
\centering
\includegraphics[width=0.830\columnwidth,trim=0 0.25bp 0 0.25bp,clip]{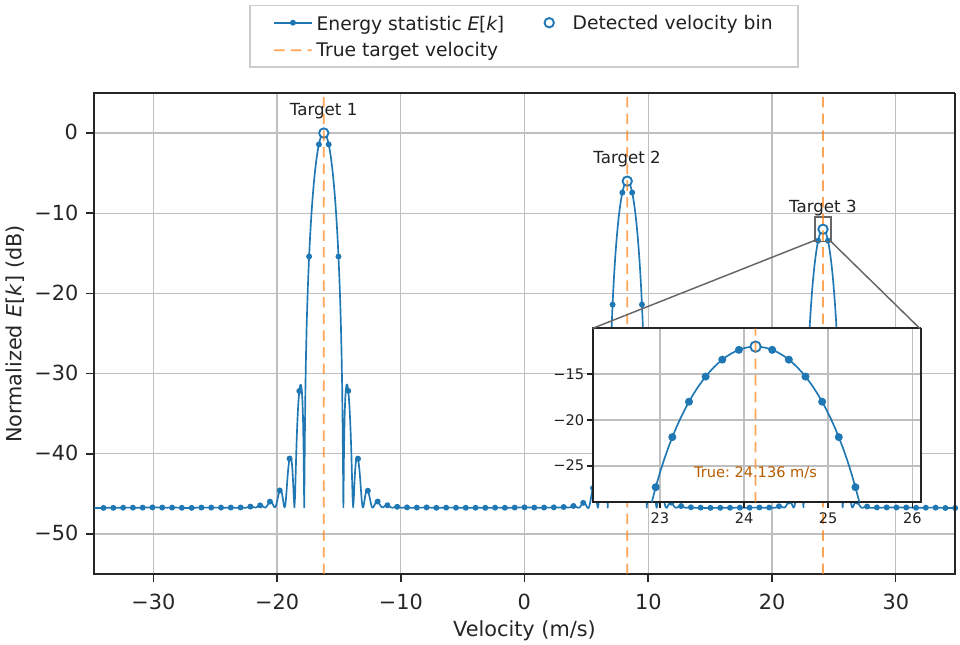}
\caption{Normalized Doppler energy statistic with a Hann window over 88 subframes and zero-padding to a 1408-point DFT. The dashed lines mark the true velocities and the hollow circles denote the selected Doppler bins.}
\label{fig:Doppler_spectrum_multitarget}
\end{figure}

By noting \eqref{eq:off_diagonal_error_ratio}, having an accurate Doppler estimation is important to prevent the appearance of ghost peak in the DOA spectrum. To improve the accuracy of Doppler estimation, a Hann window is applied over the 88 subframes, followed by zero-padding to a 1408-point DFT to refine the frequency grid, as shown in Fig. \ref{fig:Doppler_spectrum_multitarget}. The selected signed bins are $328$, $-168$, and $-488$, with $\hat f_{D,p}=\hat k_p/(1408T_r)$, corresponding to velocities of $-16.2223$, $8.3089$, and $24.1356$~m/s, respectively. These targets lie on the padded grid in this scenario. Zero-padding refines the frequency grid without improving the physical Doppler resolution. 

\begin{figure}[!t]
\centering
\includegraphics[width=0.830\columnwidth,trim=0 7.75bp 0 1.75bp,clip]{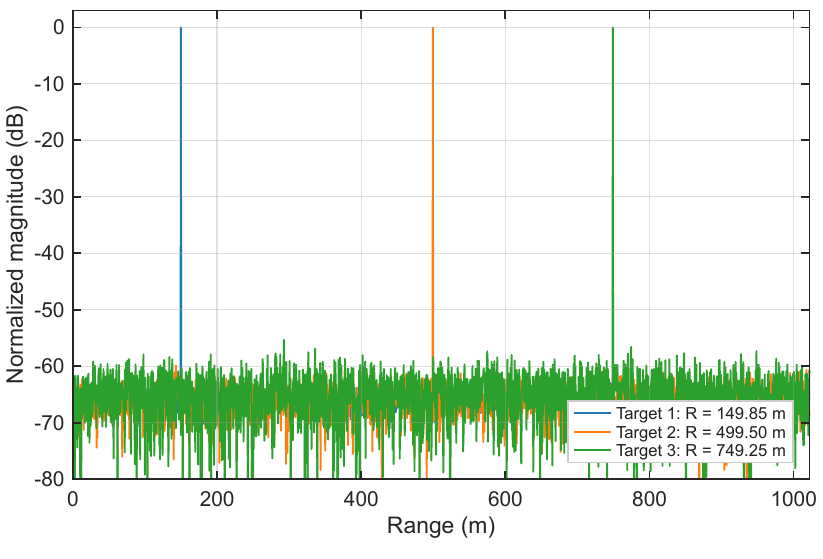}
\caption{Normalized range profiles of the three targets with Doppler-compensated STC processing.}
\label{fig:Range_profile_multitarget}
\end{figure}

\begin{figure}[!t]
\centering
\includegraphics[width=0.830\columnwidth,trim=0 0.00bp 0 0.25bp,clip]{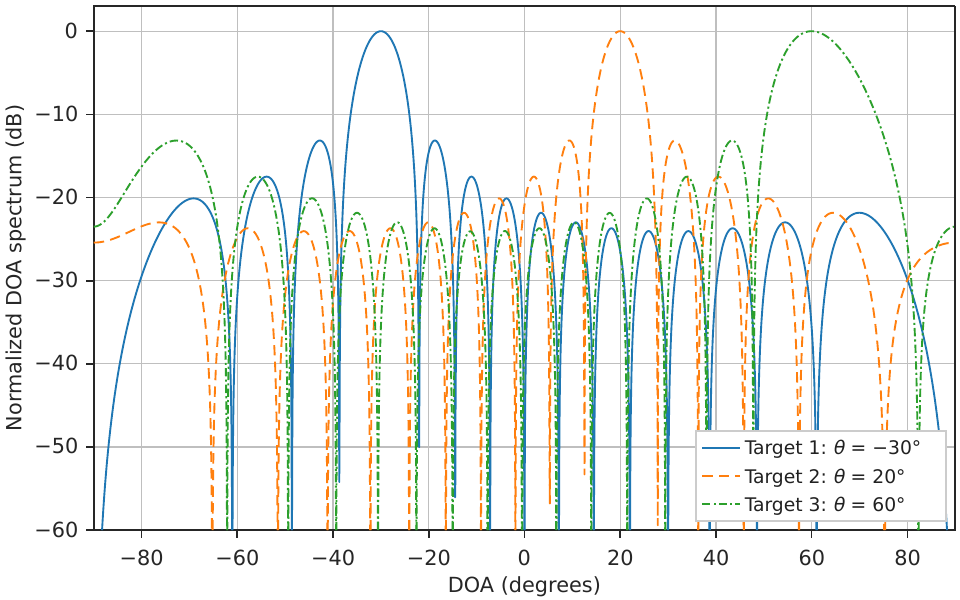}
\caption{DOA spectra of the three targets. The peaks corresponding to the DOA estimates of the targets are clearly distinguishable without the Doppler-induced ghost peaks.}
\label{fig:DOA_spectrum_multitarget}
\end{figure}

By using the refined Doppler estimation, the corresponding Doppler frequency is utilized in the fast-time and coding slot compensation for every candidate Doppler bin. Subsequently, STC decoding, range correlation with repetition accumulation, and DOA processing are performed independently for that Doppler cell. Fig. \ref{fig:Range_profile_multitarget} presents the normalized range profiles of the three detected targets. At the selected range-Doppler bin, virtual array samples are arranged for spatial processing, and Fig. \ref{fig:DOA_spectrum_multitarget} shows the DOA spectrum for each target. 
These results confirm that Doppler-bin-dependent compensation separates and accurately localizes multiple targets with distinct velocities in Doppler, range, and DOA domain.

\section{Conclusion}
This paper proposed an STC-based framework for MIMO PMCW radar that separates transmit signals before range correlation, eliminating the periodic cross-correlation terms that increase range sidelobes in conventional CDM-MIMO. The range response is consequently limited by the periodic auto-correlation of the selected sequence, allowing the same binary sequence to be reused across transmit antennas. However, moving targets introduces chip-wise Doppler distortion and inter coding slot phase shift that disturb STC orthogonality and generate DOA ghosts. It further motivates us to propose the Doppler-compensated processing, which estimates Doppler first and jointly compensates the fast-time and coding slot phases before decoding. Perfect compensation restores STC orthogonality and the code's inherent PACF. Simulation results demonstrate improved range sidelobe performance and suppression of DOA ghosts. We also extend the proposed scheme to multiple targets scenario, where we process each of detected Doppler bin separately, and the simulation results show that the distinct targets are identified accurately in the Doppler, range and DOA domain. For future work, it is worth investigating an ISAC scenario where information symbols are transmitted in the STC matrix, and the communication receiver is designed to recover these symbols.  

\bibliographystyle{IEEEtran}
\bibliography{references.bib}

\end{document}